\documentclass[english,english,pra,english,preprint,amsmath,amssymb,aps,longbibliography,showkeys,titlepage,superscriptaddress]{revtex4-2}
\usepackage{lmodern}
\usepackage{lmodern}
\usepackage[T1]{fontenc}
\usepackage[utf8]{inputenc}
\usepackage{amsmath}
\usepackage{amsthm}
\usepackage{amssymb}
\usepackage{graphicx}
\usepackage{esint}

\makeatletter

\usepackage{amsthm}\usepackage{latexsym}\usepackage{bm}\usepackage{amsfonts}
\usepackage{babel}
\usepackage{enumitem}

\usepackage[hypertexnames=false]{hyperref}
\hypersetup{
breaklinks = true,
colorlinks = true,
citecolor = {blue},
urlcolor = {blue},
linkcolor = {blue}
}

\@ifundefined{showcaptionsetup}{}{%
 \PassOptionsToPackage{caption=false}{subfig}}
\usepackage{subfig}
\makeatother

\usepackage{babel}

\begin{document}
\title{Energizing charged particles by an orbit instability in a slowly rotating
magnetic field}
\author{Eric Palmerduca}
\email{ep11@princeton.edu}

\affiliation{Princeton Plasma Physics Laboratory, Princeton University, Princeton, NJ, USA}
\affiliation{Department of Astrophysical Sciences, Princeton University, Princeton, NJ, USA}

\author{Hong Qin}
\affiliation{Princeton Plasma Physics Laboratory, Princeton University, Princeton, NJ, USA}
\affiliation{Department of Astrophysical Sciences, Princeton University, Princeton, NJ, USA}

\author{Samuel A. Cohen}
\affiliation{Princeton Plasma Physics Laboratory, Princeton University, Princeton, NJ, USA}
\affiliation{Department of Astrophysical Sciences, Princeton University, Princeton, NJ, USA}

\begin{abstract}
The stability of charged particle motion in a uniform magnetic field
with an added spatially uniform transverse rotating magnetic field
(RMF) is studied analytically. It is found that the stability diagram
of a single-particle's orbit depends critically on the chosen boundary
conditions. We show that for many boundary conditions and wide regions
in the parameter space, RMFs oscillating far below the cyclotron frequency
can cause linear instabilities in the motion which break $\mu$-invariance
and energize particles. Such energization may appear at odds with
the adiabatic invariance of $\mu$; however, adiabatic invariance
is an asymptotic result, and does not preclude such heating by magnetic
fields oscillating at slow frequencies. This mechanism may contribute
to heating in the edge plasma of field-reversed configurations (FRCs)
in rotamak-FRC experiments. Furthermore, these RMF-driven instabilities
may significantly enhance azimuthal current drive during the formation
of FRCs in such devices. 
\end{abstract}
\maketitle

\section{Introduction}

Consider a particle in a constant, uniform magnetic field $\mathbf{B}_{0}$
with cyclotron frequency $\Omega_{0}$. We examine the possibility
of destabilizing the dynamics and pumping energy into the particle
by applying an additional perpendicular spatially uniform rotating
magnetic field (RMF) $\mathbf{B}_{1}(t)$ which rotates in the transverse
plane with period $T$ much longer than the gyroperiod ($\Omega_{1}\mathrel{\dot{=}}2\pi/T\ll\Omega_{0}$).
We show that for many boundary conditions and wide regions in the
parameter space, RMFs oscillating far below the cyclotron frequency
can cause linear instabilities in the motion which break $\mu$-invariance
and energize particles. This model system of low frequency RMF has
been studied before, primarily in studies focused on finding conditions
for stable solutions \citep{Kazantsev1959, Soldatenkov1966,Kurbatov1976,Fisch1982, Rax2016, Wetering2021},
and is particularly important to current drive in field-reversed configuration
(FRC) plasmas \citep{Hugrass1982,Hugrass1983}. However, previous
treatments have not fully addressed two important aspects of the problem,
namely the roles of adiabatic invariance and boundary conditions in
the theory.

The former issue is theoretical in nature. At first glance, invariance
of the magnetic moment $\mu=W_{\perp}/B$, which is often assumed
in plasma physics, might preclude energization of the particle since
$B=|\mathbf{B}_{0}+\mathbf{B}_{1}(t)|$ is constant. Indeed, while
schemes for energizing plasmas via magnetic pumping were proposed
early in the history of plasma physics, first to explain cosmic ray
phenomena \citep{Alfven1950b}, and soon after as potential heating
mechanisms for fusion devices \citep{Spitzer1953a}, they typically
relied on breaking conditions for the adiabatic invariance of $\mu$.
In collisional magnetic pumping, dissipative, non-Hamiltonian collisions
break adiabaticity \citep{Dawson1965b}. In transit-time magnetic
pumping, particles experience rapidly changing magnetic fields when
entering and exiting a finite region of oscillating $\mathbf{B}$,
breaking the assumption that the field is slowly varying in space
\citep{Berger1954,Dawson1965b}. RMF heating of FRC plasmas has also
been explained by breaking adiabatic conditions either at magnetic
nulls \citep{Landsman} or by Speiser collisions \citep{Speiser,Glasser2002}.
In contrast, the presented model of energization by slow RMFs suggests
that $\mu$ can grow significantly without explicitly breaking adiabatic
conditions. This is possible since adiabatic invariance is an asymptotic
result, and as such it does not necessarily limit the growth of $\mu$
for fixed small values of the adiabatic parameter $\epsilon=\frac{\Omega_{1}}{\Omega_{0}}$.

This paper also answers a question of more practical import, that
of the role of boundary conditions on the stability of the particle
dynamics. The RMF $\mathbf{B}_{1}$ induces an electric field $\mathbf{E}$
which is determined in part by the boundary conditions, and is therefore
not unique. Three different sets of boundary conditions have been employed
in past studies, each leading to vastly different stability criteria
\citep{Kazantsev1959,Soldatenkov1966, Kurbatov1976,Fisch1982,Hugrass1982,Rax2016,Wetering2021}. In a study on RMF driven isotope separation, Rax and Gueroult identified that there is actually a large family of boundary conditions consistent with this problem whose stability properties remain unexplored \citep{Rax2016}. Here, we derive the complete set of consistent boundary conditions, and determine the stability criteria for a one-parameter subset of these which characterizes those configurations most relevant to laboratory plasmas. This generalized treatment reveals two important points. Firstly,
one can exhibit a large degree of control over the stability diagrams
by varying boundary conditions, and that with correctly chosen conditions,
one may be able to energize one or both species in a plasma. Such
a mechanism may provide a linearized model of heating in the edge
region in field reversed configuration (FRC) devices driven by RMFs.
In regimes in which electrons, but not ions, are energized, we show that the energy is selectively imparted into the azimuthal
motion, leading to azimuthal current drive. This mechanism may be
important in the formation of FRCs. Secondly, one can answer the question
of how sensitive the stability diagram is to perturbation in the boundary
conditions. It is shown that the stability diagrams in some previous
studies \citep{Soldatenkov1966,Fisch1982,Hugrass1983,Wetering2021} are in fact highly sensitive
to such perturbations. However, this is an exceptional case. We show
that for almost all boundary conditions the Hamiltonian structure
of the problem provides structural stability, guaranteeing that small
changes in the boundary conditions only slightly perturb the stability
diagram.

Fluid models \citep{Hugrass1981,Milroy2010} have been extensively used to study the penetration of rotating magnetic fields into FRCs. However, fluid models are not applicable when the gyro-radii of particles exceeds a moderate fraction ($\sim 1/10$) of the field curvature, a condition that exist in small $s$ (hot, kinetic) FRCs and near the minor axis (the O-point null line) of all FRCs. ($s\approx 0.3 \rho_i / r_s$, where $\rho_i$ is the ion gyro-radius at the FRC’s field maximum and $r_s$ is the separatrix radius). Supporting the latter assertion is the observation that RMF fields often do not penetrate to the minor axis of many FRCs \citep{Milroy2004}. Particle-in-cell codes \citep{Welch2010} have also been applied to the RMF/FRC problem. In addition to being self-consistent, these do address kinetic issues, though the required computational resources are extremely high. On the other hand, single particle models \citep{Glasser2022} offer a far simpler framework which allows detailed study of certain important rapid processes, independent of the complex interplay of longer duration phenomena, e.g., inductive effects and collisions, in FRCs. Applications of our model to FRCs are thus restricted to rapid processes, such as particle energization and current drive in the early stages of FRC formation.

The paper is organized as follows. The equations of motion and boundary
conditions are presented in Sec.\,\ref{sec:Equations-of-Motion},
followed by a study of the stability criteria in Sec.\,\ref{sec:Stability-Analysis}.
In Sec.\,\ref{sec:Implications} we discuss various implications
of the stability analysis, in particular, how our results compare
to previous analyses and applications to FRC devices in terms of driving
azimuthal current and heating the plasma. In Secs.\,\ref{sec:Stability-in-a}
and \ref{sec:Adiabatic-Invariance}, it is shown how energization
of charged particles by slowly oscillating fields is consistent with
the adiabatic invariance of the magnetic moment.

\section{Equations of Motion\label{sec:Equations-of-Motion}}

Consider a spatially uniform, $T$-periodic magnetic field 
\begin{equation}
\mathbf{B}(t)=\mathbf{B}_{0}\mathbf{e}_{z}+\mathbf{B}_{1}(t)\,,\label{eq:Bfield}
\end{equation}
which corresponds to the family of vector potentials 
\begin{equation}
\mathbf{A}(\mathbf{x},t)=\frac{1}{2}\mathbf{B}(t)\times\mathbf{x}+\bar{\mathbf{A}}(\mathbf{x},t)=\frac{1}{2}\mathbf{\hat{B}}(t)\mathbf{x}+\bar{\mathbf{A}}(\mathbf{x},t)\thinspace,
\end{equation}
where $\bar{\mathbf{A}}(\mathbf{x},t)$ is any curl-free vector field
and the hat-map $\hat{}:\mathbb{R}^{3}\rightarrow\mathfrak{so}(3)$ was
used on the RHS to express the cross product. We work in the Coulomb
gauge, so we have the additional constraint that $\nabla\cdot\bar{\mathbf{A}}=0$.
Therefore, $\bar{\mathbf{A}}=\nabla\psi$ where $\psi(\mathbf{x},t)$
is a solution of Laplace's equation. In the terminology of fluid dynamics,
$\mathbf{A}$ is determined up to an irrotational flow $\bar{\mathbf{A}}$.
We assume further that there is no electrostatic potential, that is,
$\phi=0$, so the electric field is purely inductive: 
\begin{equation}
\mathbf{E}(\mathbf{x},t)=-\frac{1}{c}\frac{\partial\mathbf{A}}{\partial t}=-\frac{1}{c}\big(\frac{1}{2}\mathbf{\dot{\hat{B}}}_{1}\mathbf{x}+\frac{\partial\bar{\mathbf{A}}}{\partial t}\big)\thinspace.\label{Efield}
\end{equation}
This field automatically satisfies Faraday's equation. The freedom
afforded by $\bar{\mathbf{A}}$ (or equivalently by $\psi$) can be
understood from the fact that boundary conditions have not been imposed.
Eq.\,(\ref{eq:Bfield}) describes fields that extend infinitely in
space. Such fields are nonphysical, and in reality must be coupled
to nonlinear, decaying fields for large $\mathbf{x}$. $\bar{\mathbf{A}}$
can be viewed as specifying the boundary (or matching) conditions
determined by such nonlinear fields.

We are interested in the case of spatially linear fields in which
$\mathbf{E}(\mathbf{x},t)=\mathbf{F}(t)\mathbf{x}$ for some matrix
$\mathbf{F}(t)$. Thus, $\psi$ must be a quadratic form 
\begin{equation}
\psi(\mathbf{x},t)=\frac{1}{2}\mathbf{x}^{T}\mathbf{G}(t)\mathbf{x}\thinspace,
\end{equation}
where $\mathbf{G}$ is a traceless matrix which can generally be taken
to be symmetric. We then arrive at the general form of the vector
potential 
\begin{equation}
\mathbf{A}(\mathbf{x},t)=\frac{1}{2}\big(\mathbf{\hat{B}}(t)+\mathbf{G}(t)\big)\mathbf{x}\mathrel{\dot{=}}\mathbf{\mathcal{A}}\mathbf{x}.\label{eq:vec_pot}
\end{equation}
The dynamics of a particle with charge $q$ and mass $m$ in this
field are described by the Hamiltonian 
\begin{equation}
H(\mathbf{x},\mathbf{p})=\frac{1}{2m}\Big(\mathbf{p}-\frac{q}{c}\mathbf{A}\Big)^{2}\thinspace,
\end{equation}
where $\mathbf{p}$ is the canonical momentum 
\begin{equation}
\mathbf{p}=m\mathbf{v}+\frac{q}{c}\mathbf{A}.
\end{equation}
Hamilton's equations then give the canonical equations of motion 
\begin{align}
 & \mathbf{\dot{x}}=\frac{\partial H}{\partial\mathbf{p}}=\frac{1}{m}\big(\mathbf{p}-\frac{q}{c}\mathbf{\mathcal{A}}\mathbf{x}\big)\thinspace,\\
 & \mathbf{\dot{p}}=-\frac{\partial H}{\partial\mathbf{x}}=\frac{q}{mc}\mathbf{\mathcal{A}}^{T}\big(\mathbf{p}-\frac{q}{c}\mathcal{A}\mathbf{x}\big)\thinspace.
\end{align}
We normalize time using the background gyrofrequency $\Omega_{0}\mathrel{\dot{=}}\frac{qB_{0}}{mc}$:
\begin{equation}
\tau=\Omega_{0}t,\;\;\mathbf{\tilde{x}}(\tau)=\mathbf{x}(t),\;\;\mathbf{\tilde{p}}(\tau)=\frac{\mathbf{p}(t)}{m\Omega_{0}},\;\;\mathbf{\tilde{B}}(\tau)=\frac{\mathbf{B}(t)}{B_{0}},\;\;\ensuremath{\tilde{\mathbf{\mathcal{A}}}}=\frac{\mathbf{\mathcal{A}}}{B_{0}},\;\;\tilde{T}=\Omega_{0}T
\end{equation}
to obtain the form 
\begin{equation}
\begin{pmatrix}\mathbf{\dot{\tilde{x}}}\\
\mathbf{\dot{\tilde{p}}}
\end{pmatrix}=\begin{pmatrix}-\tilde{\mathbf{\mathcal{A}}} & \mathbf{I}\\
-\tilde{\mathbf{\mathcal{A}}}^{T}\mathbf{\tilde{\mathbf{\mathcal{A}}}} & \tilde{\mathbf{\mathcal{A}}}^{T}
\end{pmatrix}\begin{pmatrix}\mathbf{\tilde{x}}\\
\mathbf{\tilde{p}}
\end{pmatrix}\mathrel{\dot{=}}\mathbf{M}(\tau)\begin{pmatrix}\mathbf{\tilde{x}}\\
\mathbf{\tilde{p}}
\end{pmatrix};\;\;\mathbf{M}(\tau+\tilde{T})=\mathbf{M}(\tau)\thinspace.\label{EOM}
\end{equation}
This is a linear non-autonomous system with periodic coefficients,
and its solution can be formally written as 
\begin{equation}
\begin{pmatrix}\mathbf{\tilde{x}}\\
\mathbf{\tilde{p}}
\end{pmatrix}=\mathbf{P}(\tau)\begin{pmatrix}\mathbf{\tilde{x}}(0)\\
\mathbf{\tilde{p}}(0)
\end{pmatrix}\thinspace,
\end{equation}
where the solution map $\mathbf{P}(t)$ is determined by 
\begin{equation}
\mathbf{\dot{P}}=\mathbf{M}(\tau)\mathbf{P},\;\;\mathbf{P}(0)=\mathbf{I}\thinspace,\label{principal}
\end{equation}
where $\mathbf{I}$ is the identity. The stability of the system is
determined by the eigenvalues of the one-period solution map, also
known as the monodromy matrix, $\mathbf{P}(\tilde{T})$. In the general
case, $\mathbf{P}(\tilde{T})$ can only be obtained numerically. In
this study we will specialize to a particular family of vector potentials
for which the stability of Eq.\,(\ref{EOM}) can be studied via analytic
means. We note that this system has a Hamiltonian structure and therefore
may be amenable to analysis by a recently developed generalized Floquet
theory for non-periodic Hamiltonian systems \citep{Qin2019LH}. This
possibility will be explored in future work.

Consider the rotating magnetic field (RMF) 
\begin{equation}
\mathbf{\tilde{B}}(\tau)=\mathbf{e}_{z}+\beta\Big(\cos\epsilon\tau\:\mathbf{e}_{x}+\sin\epsilon\tau\,\mathbf{e}_{y}\Big)\thinspace,\label{eq:RMF_field}
\end{equation}
where $\epsilon\mathrel{\dot{=}}\alpha^{-1}\mathrel{\dot{=}}\frac{\Omega_{1}}{\Omega_{0}}$,
$\beta=\frac{B_{1}}{B_{0}}$, and $\Omega_{1}=\frac{2\pi}{T}$. Such fields have been employed in multiple plasma physics applications, such as in simplified models of the applied fields in Rotamak-FRCs \citep{Fisch1982, Wetering2021, Hugrass1982, Hugrass1983} and in an RMF-driven plasma separation concept \citep{Rax2016}. In the slowly rotating limit $|\epsilon|\rightarrow 0$, it can be used as a model system to study the interplay between adiabatic invariance and particle energization. The
orientation of the rotation with respect to the particle gyration
in the background field is specified by the sign of $\epsilon$ (or
equivalently $\alpha$) with $\epsilon>0$ and $\epsilon<0$ corresponding
to counter- and co-rotating fields, respectively. In this case, Eq.\,(\ref{eq:vec_pot})
gives 
\begin{equation}
\mathbf{\mathcal{\tilde{A}}}=\frac{\hat{\mathbf{e}}_{z}}{2}+\frac{\beta}{2}\begin{pmatrix}g_{11}(\tau) & g_{12}(\tau) & \sin\epsilon\tau+g_{13}(\tau)\\
g_{12}(\tau) & g_{22}(\tau) & -\cos\epsilon\tau+g_{23}(\tau)\\
-\sin\epsilon\tau+g_{13}(\tau) & \cos\epsilon\tau+g_{23}(\tau) & -(g_{11}(\tau)+g_{22}(\tau))
\end{pmatrix}\thinspace.\label{eq:A_general}
\end{equation}
In general, the $g_{ij}(\tau)$ are arbitrary functions of time determined
by the boundary conditions. However, typical conditions of theoretical
and experimental interest substantially constrain the $g_{ij}(\tau)$.
We consider the case in which the boundary conditions have the same
driving frequency and are in phase with the RMF (and with no higher
harmonics): 
\begin{align}
g_{13}(\tau)=b_{13}\sin\epsilon\tau\thinspace,\\
g_{23}(\tau)=a_{23}\cos\epsilon\tau\thinspace,\\
g_{ij}(\tau)=a_{ij}\cos\epsilon\tau+b_{ij}\sin\epsilon\tau; & \;\;i\leq j,\;j\in\{1,2\}\thinspace,
\end{align}
where the $a_{ij}$ and $a_{ij}$ are constants. We also assume that
the boundary conditions are rotationally symmetric with respect to
the RMF. That is, we assume the cylindrical components of $\mathbf{E}$
satisfy 
\begin{alignat}{1}
E_{r}(r,\phi+\epsilon\tau',z,\tau+\tau')=E_{r}(r,\phi,z,\tau)\thinspace,\label{eq:E_r}\\
E_{\phi}(r,\phi+\epsilon\tau',z,\tau+\tau')=E_{\phi}(r,\phi,z,\tau)\thinspace,\label{eq:E_phi}\\
E_{z}(r,\phi+\epsilon\tau',z,\tau+\tau')=E_{z}(r,\phi,z,\tau)\thinspace.\label{eq:E_z}
\end{alignat}
These conditions essentially assert that the only preferred direction
in the xy-plane is that specified by the instantaneous direction of
the RMF. We can eliminate many of the $A_{ij}$ and $B_{ij}$ using
these conditions. For example, with $\tau=0$ and $\phi=0$, condition
(\ref{eq:E_z}) can be expressed as
\begin{equation}
-\frac{B_{1}\Omega_{1}}{2c}\Big[r(a_{23}+b_{13})(1-\cos2\epsilon\tau)+2z(a_{11}+b_{22})(-1+\cos\epsilon\tau)-2z(a_{11}+b_{22})\sin\epsilon\tau\Big]=0.
\end{equation}
From the functional independence of the involved trig functions, it
follows that $a_{23}=-b_{13},$ $a_{11}=-a_{22}$, and $b_{11}=-b_{22}$.
Applying similar reasoning to Eqs. (\ref{eq:E_r}) and (\ref{eq:E_phi})
shows that $g_{11}$, $g_{12}$, and $g_{22}$ must vanish, leaving
a single free parameter $p\mathrel{\dot{=}}a_{23}=-b_{13}$ to specify
boundary conditions. The total electric field is then 
\begin{equation}
\mathbf{E}(\mathbf{x},\tau)=\frac{B_{1}\Omega_{1}}{2c}\big[-z(1-p)\cos(\epsilon\tau-\phi)\mathbf{e}_{r}-z(1-p)\sin(\epsilon\tau-\phi)\mathbf{e}_{\phi}+r(1+p)\cos(\epsilon\tau-\phi)\mathbf{e}_{z}\big]\label{eq:E_field}
\end{equation}
which clearly satisfies conditions (\ref{eq:E_r}-\ref{eq:E_z}).
Note that $p=1$ correspond to the case when $\mathbf{E}$ is purely
in the z-direction and $p=-1$ to the case when it is only in the
transverse plane. We thus arrive at the general form of $\tilde{\mathbf{\mathcal{A}}}$:

\begin{equation}
\tilde{\mathbf{\mathcal{A}}}=\frac{\tilde B_{0}}{2}\hat{\mathbf{e}}_z+\frac{\beta}{2}\begin{pmatrix}0 & 0 & (1-p)\sin\epsilon\tau\\
0 & 0 & -(1-p)\cos\epsilon\tau\\
-(1+p)\sin\epsilon\tau & (1+p)\cos\epsilon\tau & 0
\end{pmatrix}.\label{eq:A_matrix}
\end{equation}
We have made these assumptions both because they are typical of laboratory
applications of RMFs as well as because they allow a vastly simplified
analytical treatment by moving into a rotating frame. That said, we
must take care that the results of the following stability analysis
are robust against small perturbations away from these specialized
boundary conditions. This point is addressed in Sec.\,\ref{subsec:Geometric-Stability}
where it is shown that the Hamiltonian structure guarantees such robustness
for almost all values of $p$.

We briefly remark on the question of determining $p$ in practice.
A choice of $\bar{\mathbf{A}}$, or equivalently of $p$, mathematically
represents a boundary condition on $\mathbf{E}$. In practical scenarios,
the uniform rotating $\mathbf{B}$ and linear rotating $\mathbf{E}$
configuration assumed will only hold locally. Thus, by linearizing
$\mathbf{E}$ in a region of nearly uniform rotating $\mathbf{B}$,
an equation of the form (\ref{eq:E_field}) is obtained, from which
one can read off $p$.

The explicit time dependence in Eq.\,(\ref{eq:A_matrix}) can be
eliminated by transforming into the coordinate system $(\mathbf{x'},\mathbf{p'})$
rotating with the RMF: 
\begin{align}
\mathbf{x}' & =\mathbf{R}(\epsilon\tau)\tilde{\mathbf{x}}\label{eq:x_trans}\\
\mathbf{p}' & =\mathbf{R}(\epsilon\tau)\tilde{\mathbf{p}}\label{eq:p_trans}
\end{align}
where $\mathbf{R}(\theta)$ is the xy-rotation matrix 
\begin{equation}
\mathbf{R}(\theta)=\begin{pmatrix}\cos\theta & \sin\theta & 0\\
-\sin\theta & \cos\theta & 0\\
0 & 0 & 1
\end{pmatrix}.
\end{equation}
We note that this is a canonical transformation since it is the cotangent
lift of the point-transformation given by Eq. (\ref{eq:x_trans}).
The transformed equation of motion is then

\begin{equation}
\begin{pmatrix}\mathbf{\dot{x}}'\\
\mathbf{\dot{p}}'
\end{pmatrix}=\mathbf{M}'\begin{pmatrix}\mathbf{x}'\\
\mathbf{p}'
\end{pmatrix}\thinspace,\label{eq:EOM}
\end{equation}
where 
\begin{equation}
\mathbf{M'}=\begin{pmatrix}0 & \frac{1}{2}+\epsilon & 0 & 1 & 0 & 0\\
-\frac{1}{2}-\epsilon & 0 & \frac{\beta}{2}(1-p) & 0 & 1 & 0\\
0 & -\frac{\beta}{2}(1+p) & 0 & 0 & 0 & 1\\
-\frac{1}{4} & 0 & \frac{\beta}{4}(1-p) & 0 & \frac{1}{2}+\epsilon & 0\\
0 & -\frac{1}{4}[1+\beta^{2}(1+p)^{2}] & 0 & -\frac{1}{2}-\epsilon & 0 & \frac{\beta}{2}(1+p)\\
\frac{\beta}{4}(1-p) & 0 & -\frac{\beta^{2}}{4}(1-p)^{2} & 0 & -\frac{\beta}{2}(1-p) & 0
\end{pmatrix}.\label{eq:M_prime}
\end{equation}

\section{Stability Analysis\label{sec:Stability-Analysis}}

Since this transformed matrix $\mathbf{M}'$ is time-independent,
its eigenvalues $\lambda_{i}$ determine the stability. In particular,
this matrix is Hamiltonian (i.e. $\mathbf{M}'\in\mathfrak{sp}(6,\mathbb{R})$),
so the dynamics are stable if and only if all of its eigenvalues are
imaginary and semisimple. In the unstable case, the growth rate of
the instability is 
\begin{equation}
\gamma\mathrel{\dot{=}}\max\text{Re}\lambda_{i}.
\end{equation}
Since $\mathbf{M}'$ is Hamiltonian, its characteristic polynomial
is even: 
\begin{align}
\lambda^{6} & +\lambda^{4}(1+\beta^{2}+2\epsilon+2\epsilon^{2})\nonumber \\
 & +\lambda^{2}\epsilon\big[-p\beta^{2}+\epsilon-\frac{1}{4}\epsilon\beta^{2}(p^{2}+6p-3)+2\epsilon^{2}+\epsilon^{3}\big]+\frac{\epsilon^{3}}{4}(1+\epsilon)(1-p)^{2}\beta^{2}.\label{eq:char_poly_lam}
\end{align}
Let $u=\lambda^{2}$, so that the eigenvalues are determined by the
roots of the cubic polynomial: 
\begin{align}
u^{3}+u^{2}(1+\beta^{2}+2\epsilon+2\epsilon^{2})+u\epsilon\big[-p\beta^{2}+\epsilon-\frac{1}{4}\epsilon\beta^{2}(p^{2} & +6p-3)+2\epsilon^{2}+\epsilon^{3}\big]\nonumber \\
 & +\frac{\epsilon^{3}}{4}(1+\epsilon)(1-p)^{2}\beta^{2}=0.\label{eq:char_poly_u}
\end{align}
In particular, the system is unstable unless all roots $u_{i}$ are
real and negative, with special care being taken in the case of repeated
roots. It is possible to write down explicit expressions for these
roots, but they are extremely unwieldy. Furthermore, we are interested
not so much in the explicit solutions themselves, but in the stability
boundaries in parameter space. These are determined in part by the
zeros of the discriminant of the cubic equation, which, in this case,
is a 12th order polynomial in the parameters. A general exact analytical
treatment is therefore not possible except for special choices of
the parameters. However, we can derive all of the notable features
of the stability diagram using perturbation theory.

\subsection{Slowly rotating limit\label{subsec:Slowly-rotating-limit}}

\begin{figure}
\subfloat[]{\includegraphics[scale=0.5]{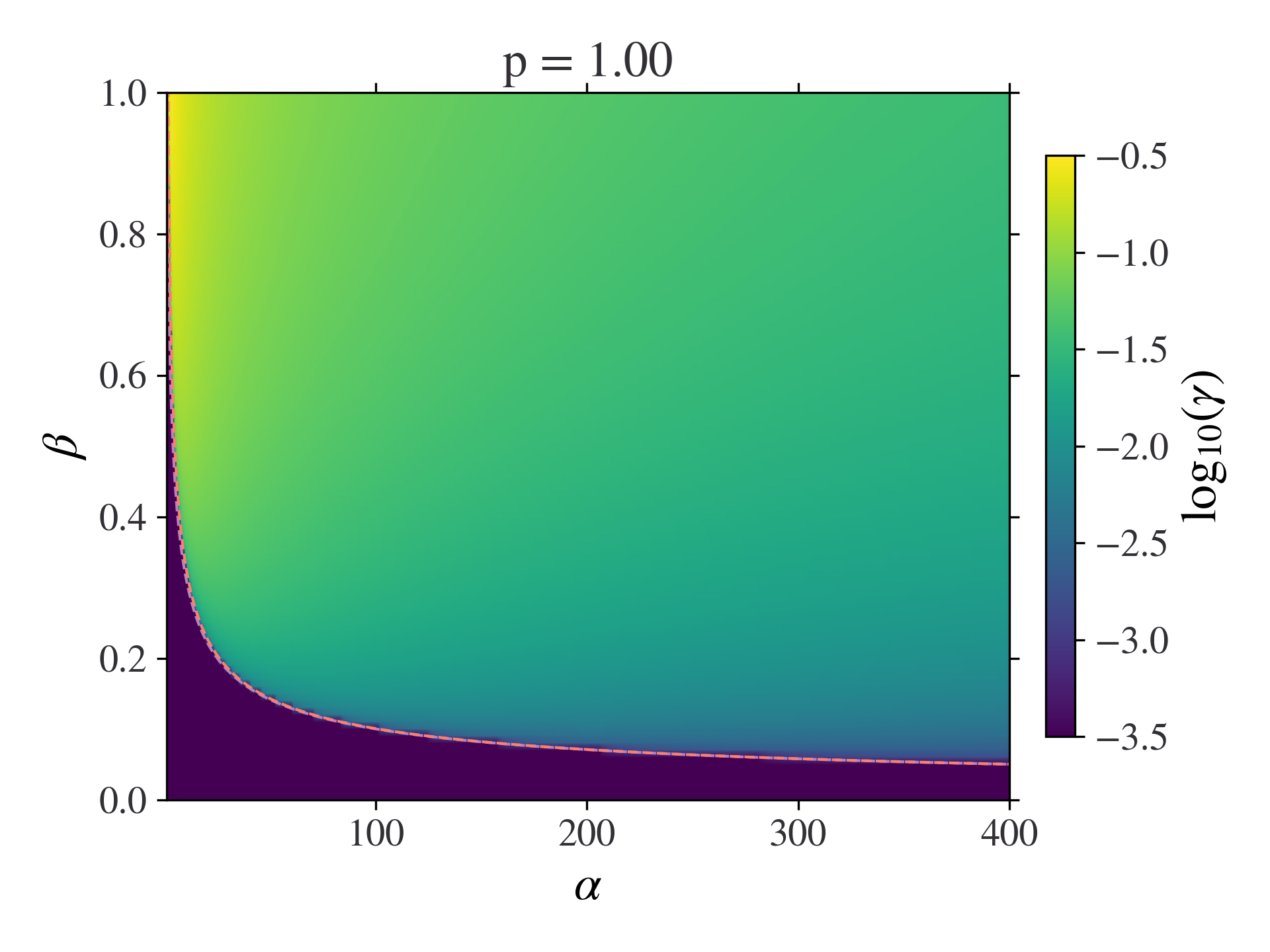}

}\subfloat[]{\includegraphics[scale=0.5]{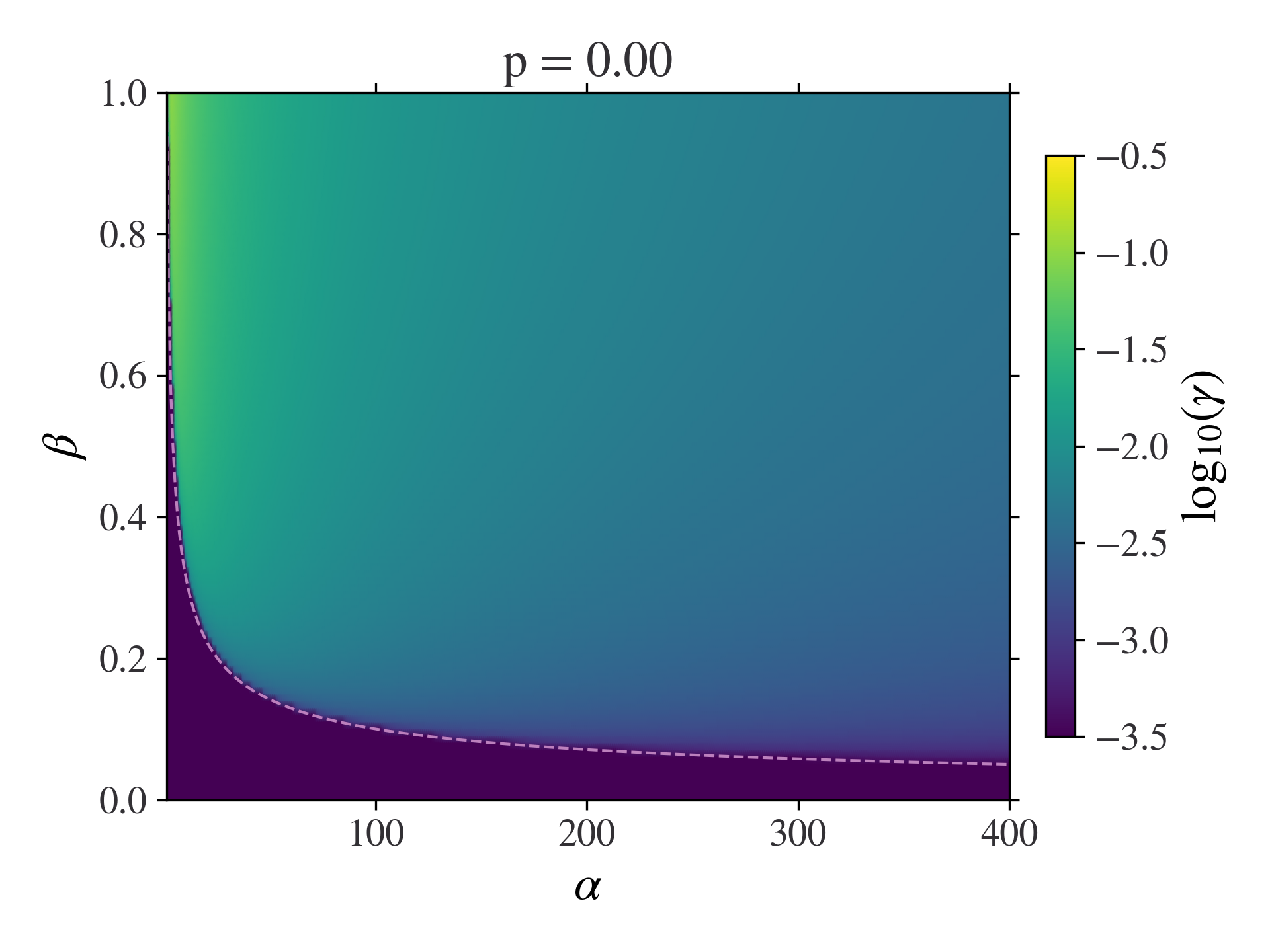}

}

\subfloat[]{\includegraphics[scale=0.5]{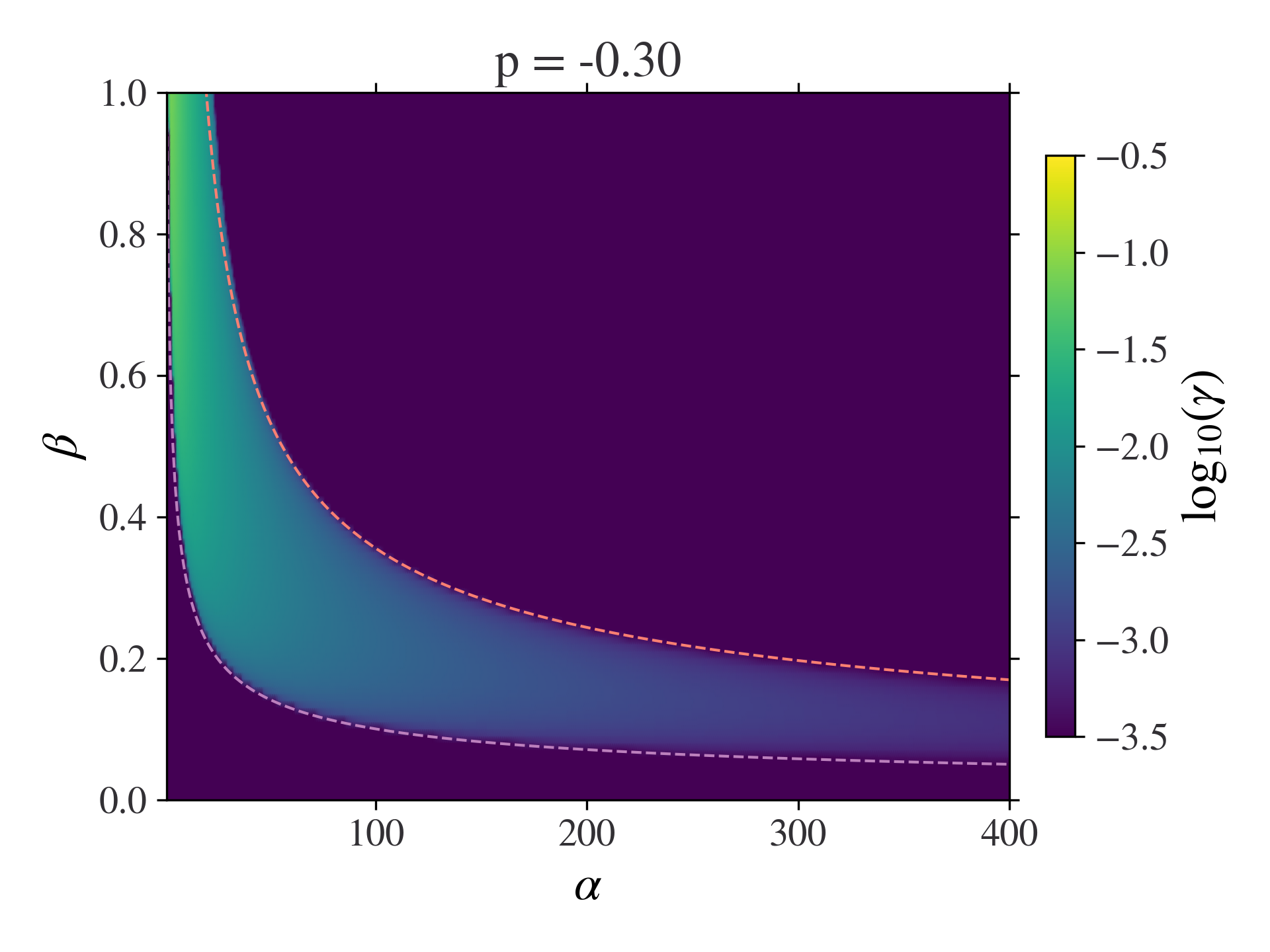}

}\subfloat[]{\includegraphics[scale=0.5]{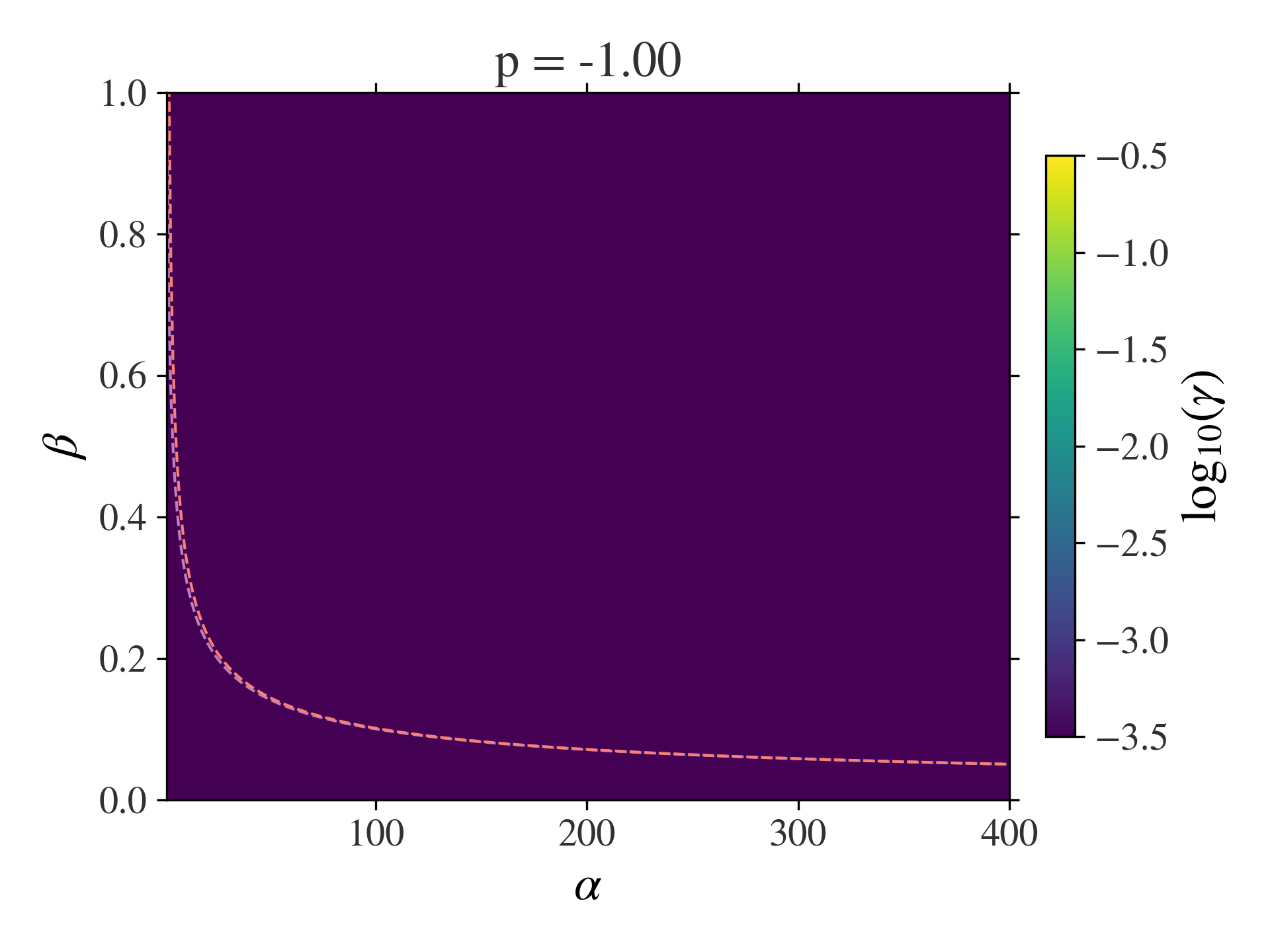}

}

\caption{Stability diagrams of a particle in a counter-rotating magnetic field
for the following boundary conditions: (a) $p=1$, (b) $p=0$, (c)
$p=-0.3$, and (d) $p=-1$. The log of the growth rate $\gamma$ is
shown as a function of normalized RMF strength $\beta=B_{1}/B_{0}$
and inverse RMF frequency $\alpha=\Omega_{0}/\Omega_{1}$. $p=1$
and $p=-1$ correspond to $\mathbf{E}\parallel\mathbf{e}_{z}$ and
$\mathbf{E}\perp\mathbf{e}_{z}$, respectively.}
\label{fig:counter_stability} 
\end{figure}

\begin{figure}
\subfloat[]{\includegraphics[scale=0.5]{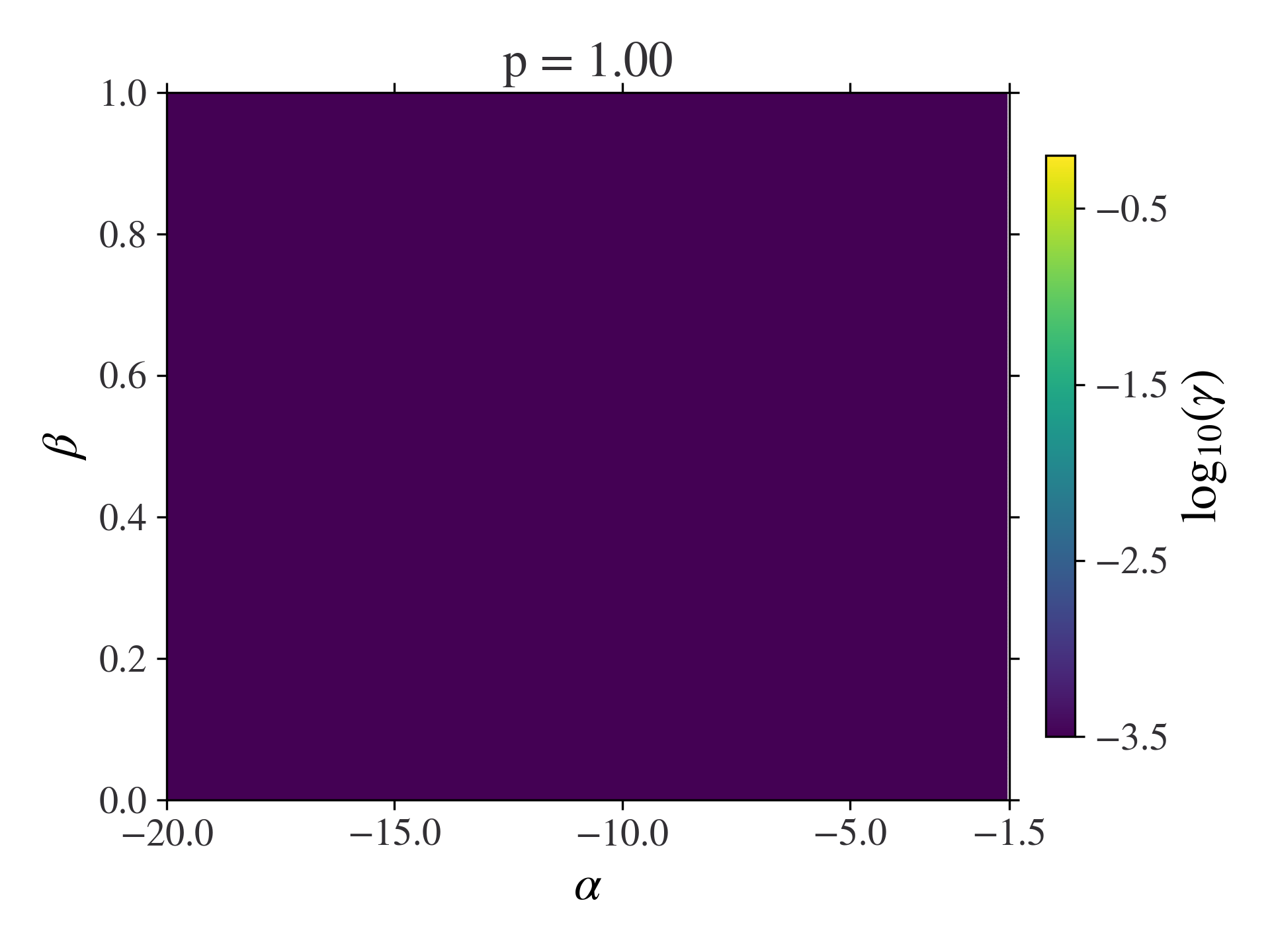}

}\subfloat[]{\includegraphics[scale=0.5]{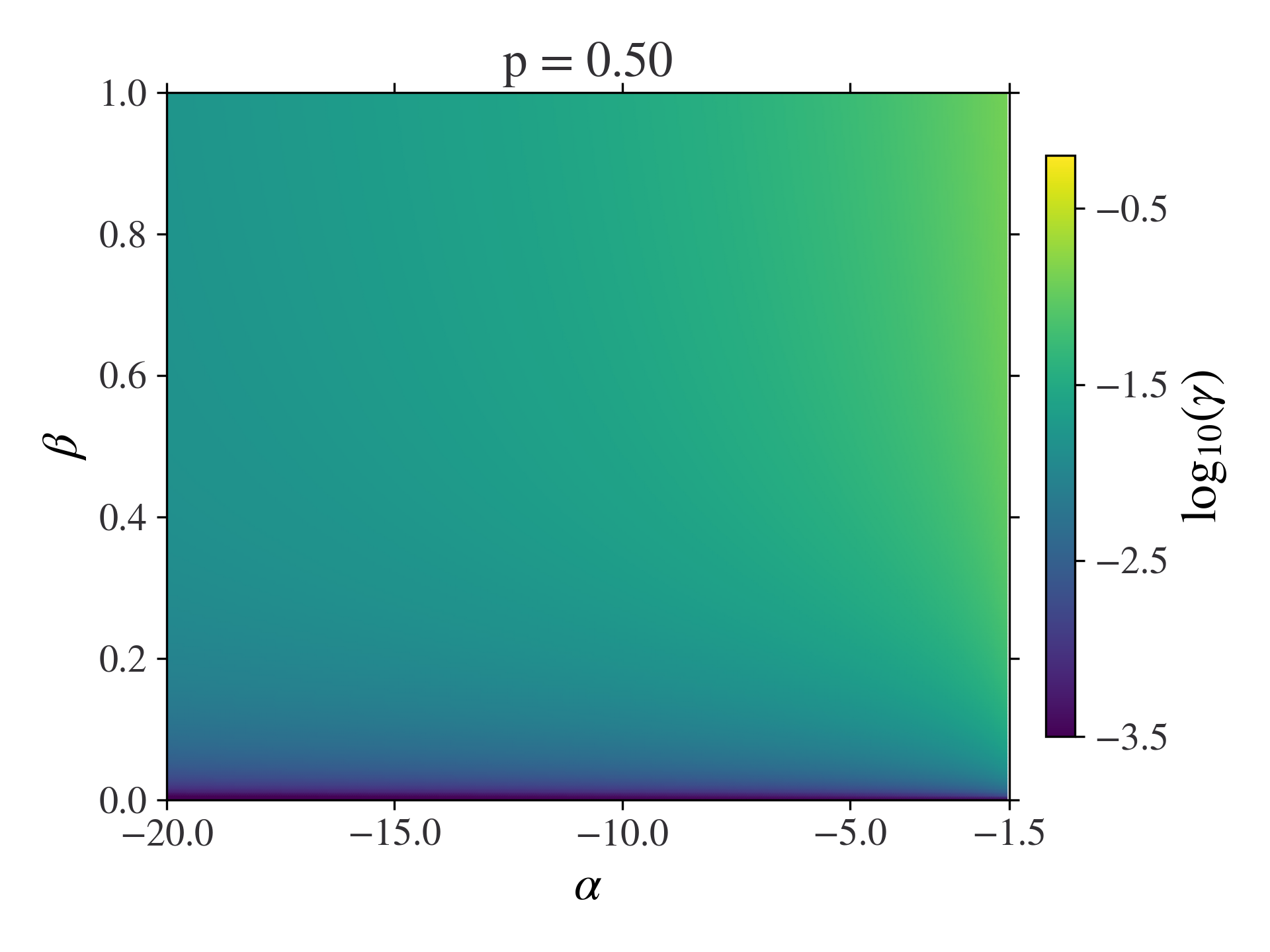}

}

\subfloat[]{\includegraphics[scale=0.5]{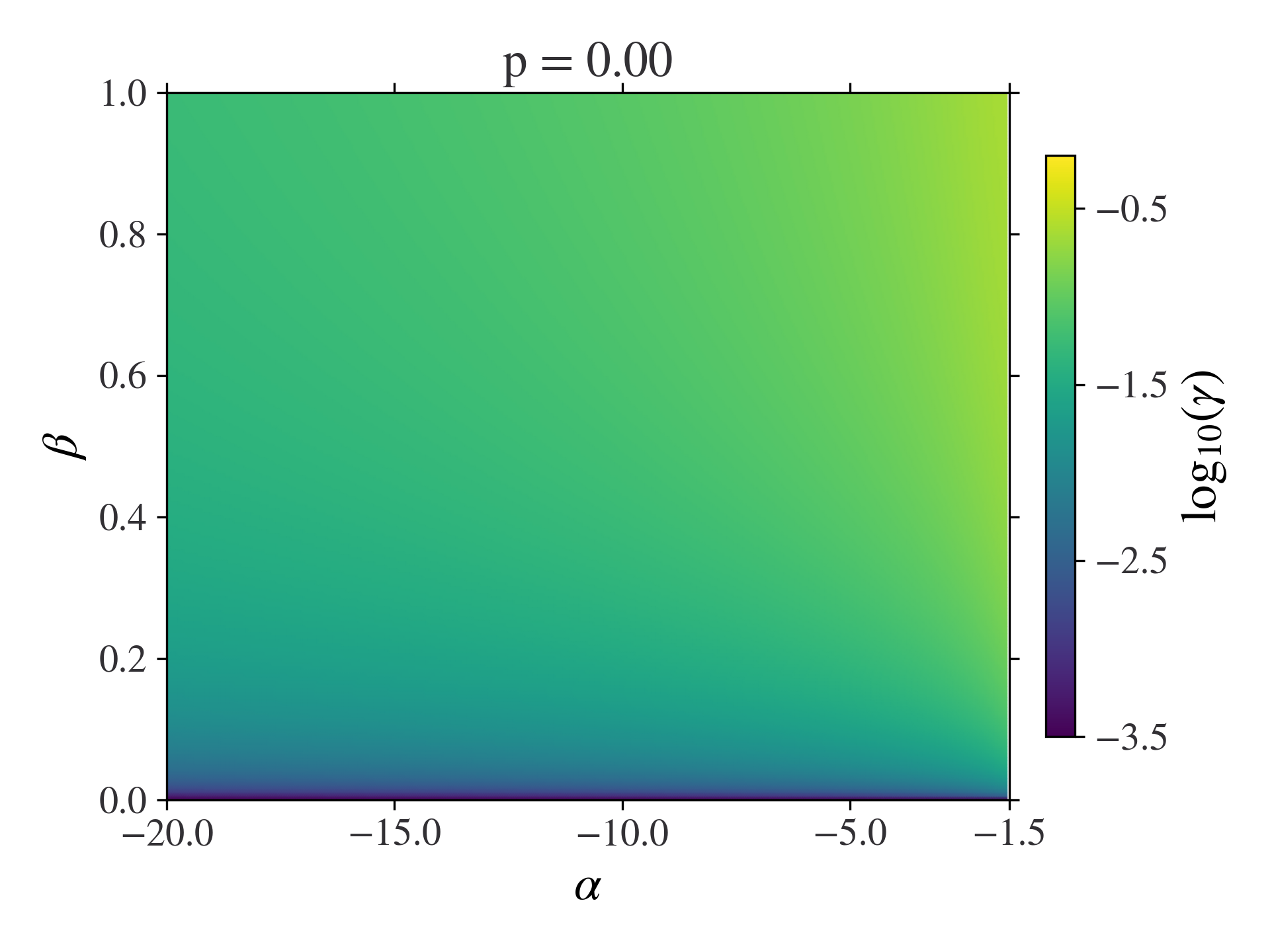}

}\subfloat[]{\includegraphics[scale=0.5]{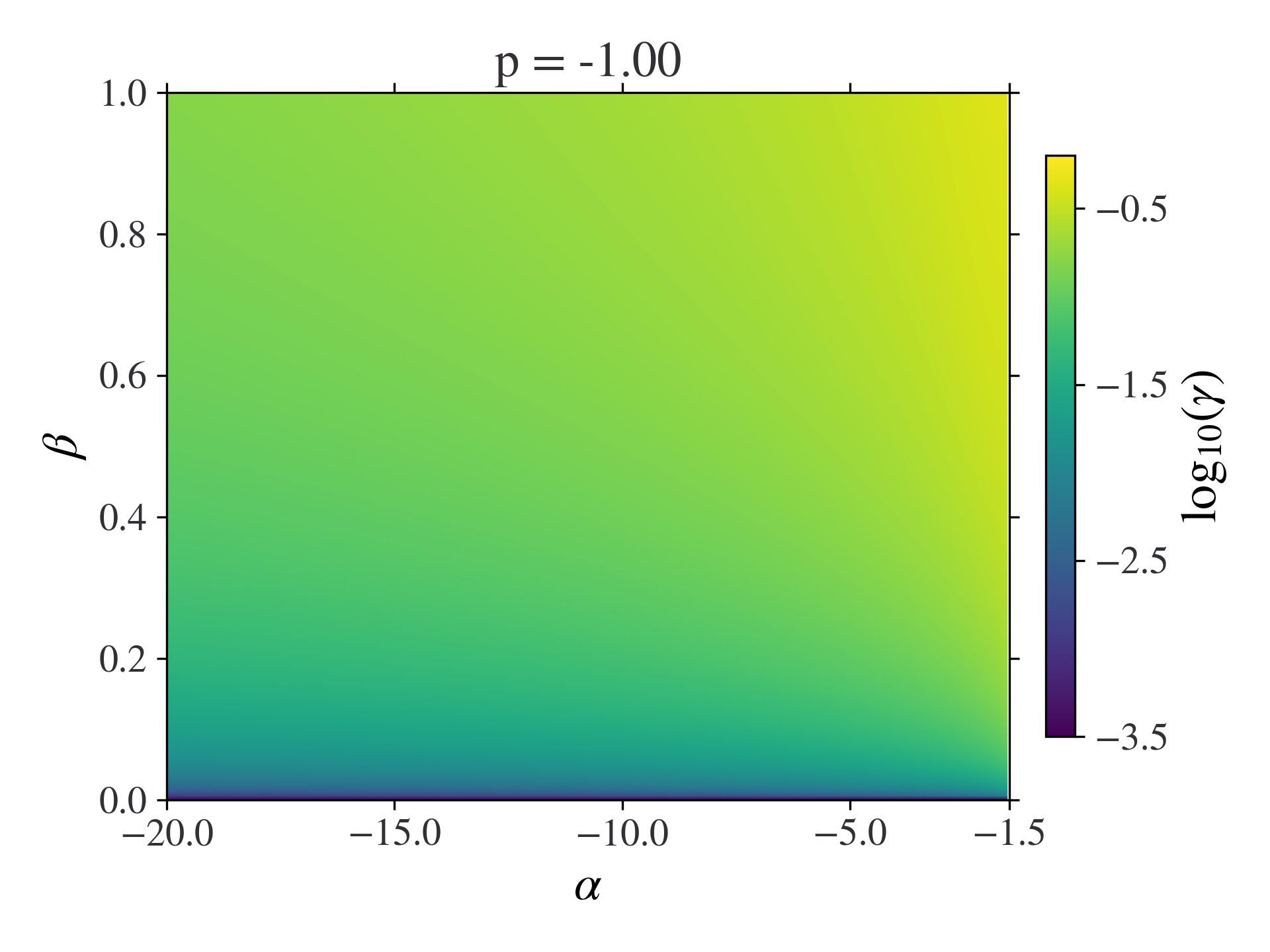}

}

\caption{Stability diagrams of a particle in a co-rotating magnetic field for
the following boundary conditions: (a) $p=1$, (b) $p=0.5$, (c) $p=0$,
and (d) $p=-1$. The log of the growth rate $\gamma$ is shown as
a function of normalized RMF strength $\beta=B_{1}/B_{0}$ and inverse
RMF frequency $\alpha=\Omega_{0}/\Omega_{1}$. $p=1$ and $p=-1$
correspond to $\mathbf{E}\parallel\mathbf{e}_{z}$ and $\mathbf{E}\perp\mathbf{e}_{z}$,
respectively.}
\label{fig:co_stability} 
\end{figure}

\begin{figure}
\subfloat[]{\includegraphics[scale=0.5]{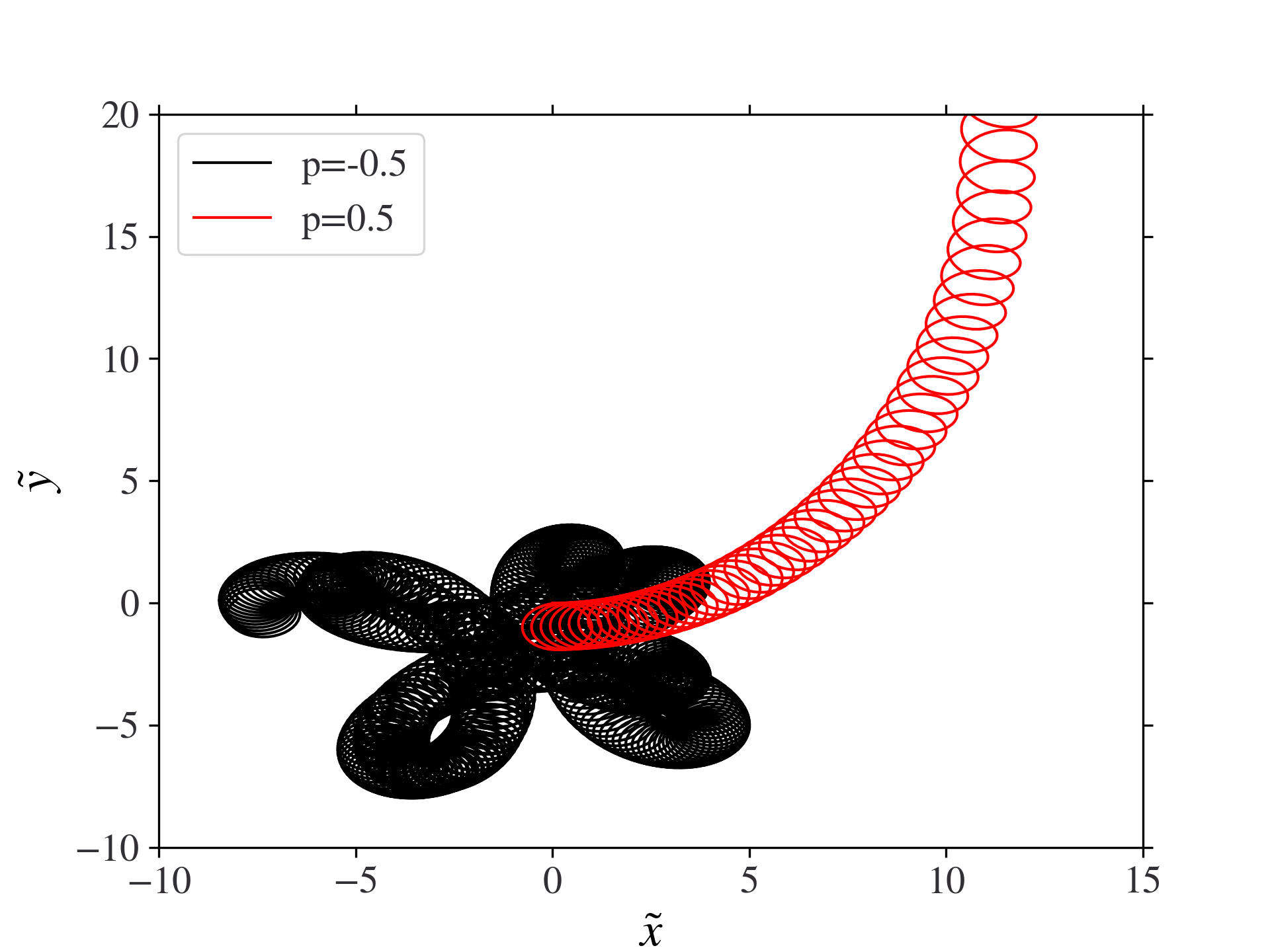}

}\subfloat[]{\includegraphics[scale=0.5]{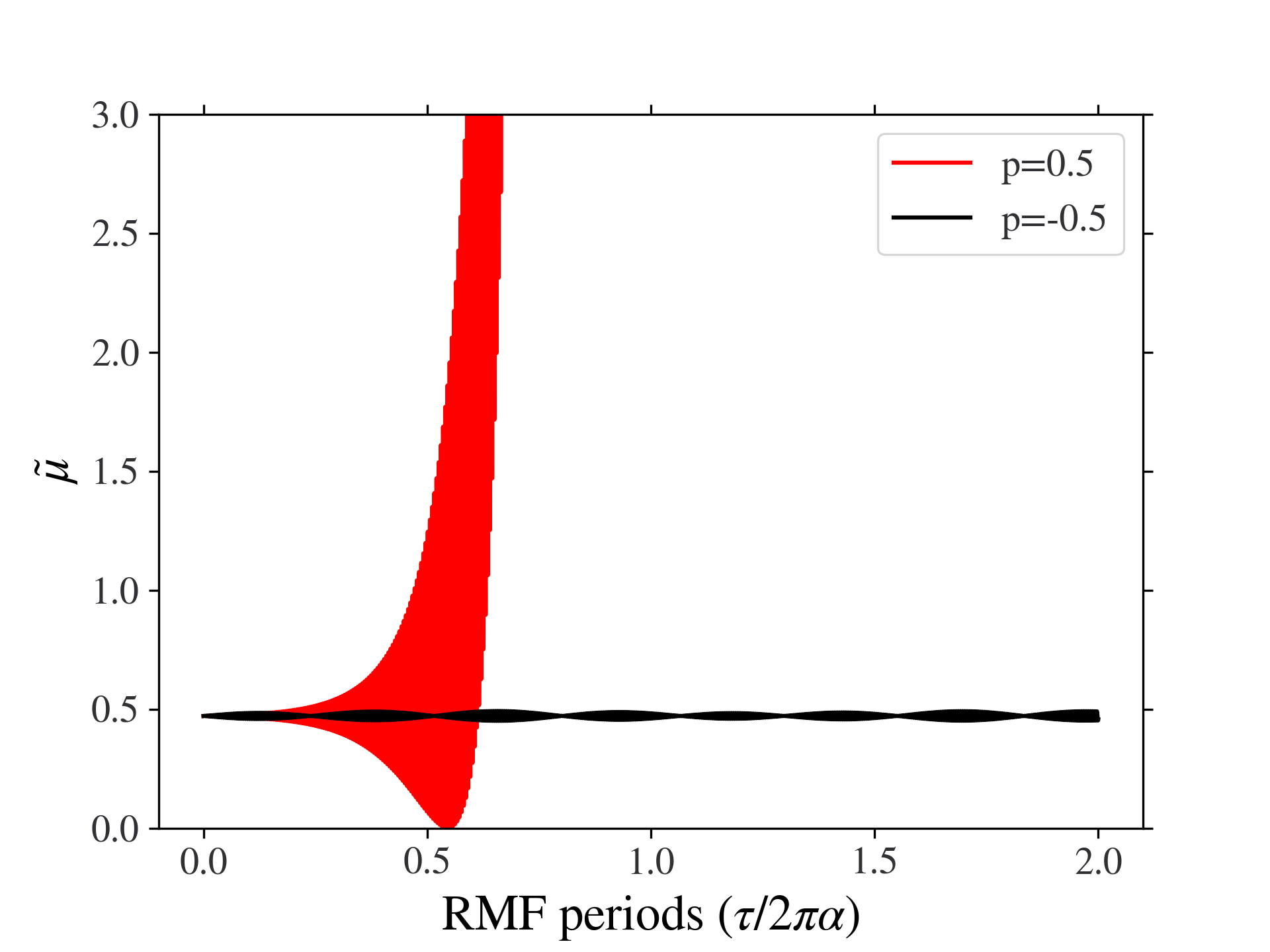}

}

\caption{(a) $\tilde{x}\tilde{y}$-projection of the trajectory and (b) magnetic
moment $\tilde{\mu}$ of a test particle in a slow counter-RMF with
$\alpha=200$, $\beta=0.2$ over 2 RMF periods (400 gyroperiods) for
two different boundary conditions: $p=0.5$ (red) and $p=-0.5$ (black).
The particle is initialized at $\mathbf{\tilde{x}=0}$ and $\mathbf{\tilde{p}}=(1,0,0).$
The stability of the particle orbit depends on the sign of $p$.}
\label{fig:large_alpha_trajectories} 
\end{figure}

The stability diagrams for the counter-rotating ($\alpha=\epsilon^{-1}>0$)
and co-rotating ($\alpha<0$) cases are shown in Figs.\,\ref{fig:counter_stability}
and \ref{fig:co_stability} for different p values between -1 and
1. We look first at the limit of slowly rotating fields $|\alpha|\gg1.$
In this case, the leading order solutions of Eq.\,(\ref{eq:char_poly_u})
are 
\begin{align}
u_{1} & \sim\frac{(p-1)^{2}}{4p\alpha^{2}}+O(\alpha^{-3})\nonumber \\
u_{2} & \sim\frac{p\beta^{2}}{\alpha(1+\beta^{2})}+O(\alpha^{-2}),\quad(|\alpha|\gg1)\label{eq:slow_sol}\\
u_{3} & \sim-(1+\beta^{2})+O(\alpha^{-1})\nonumber 
\end{align}
assuming $p\neq0,1$. In the counter-rotating ($\alpha>0)$ case,
the dynamics are stable when $p<0$ and unstable when $p>0$ with
$\gamma\sim\alpha^{-1/2}$. This difference in stability can be seen
in numerically calculated particle trajectories and the corresponding
magnetic moment evolutions shown in Fig.\,\ref{fig:large_alpha_trajectories}.
These trajectories were computed with $\alpha=200$, $\beta=0.2$,
and $p=\pm0.5$. On the other hand, in the co-rotating case either
$u_{1}$ or $u_{2}$ is always positive, so the system is always unstable
in the large $\alpha$ limit. When $p>0$, the instability rate goes
as $\gamma\sim|\alpha|^{-1/2}$ while for $p<0$, $\gamma\sim|\alpha|^{-1}$.

When $p=0$, $u_{1}$ and $u_{2}$ are given by 
\begin{equation}
u_{1,2}\sim\pm\frac{i\beta}{2\alpha^{3/2}\sqrt{1+\beta^{2}}}+O(\alpha^{-2})\label{eq:p0}
\end{equation}
and therefore the $p=0$ case is always unstable for $|\alpha|\gg1$
with $\gamma\sim|\alpha|^{-3/4}$.

\begin{figure}
\subfloat[]{\includegraphics[scale=0.5]{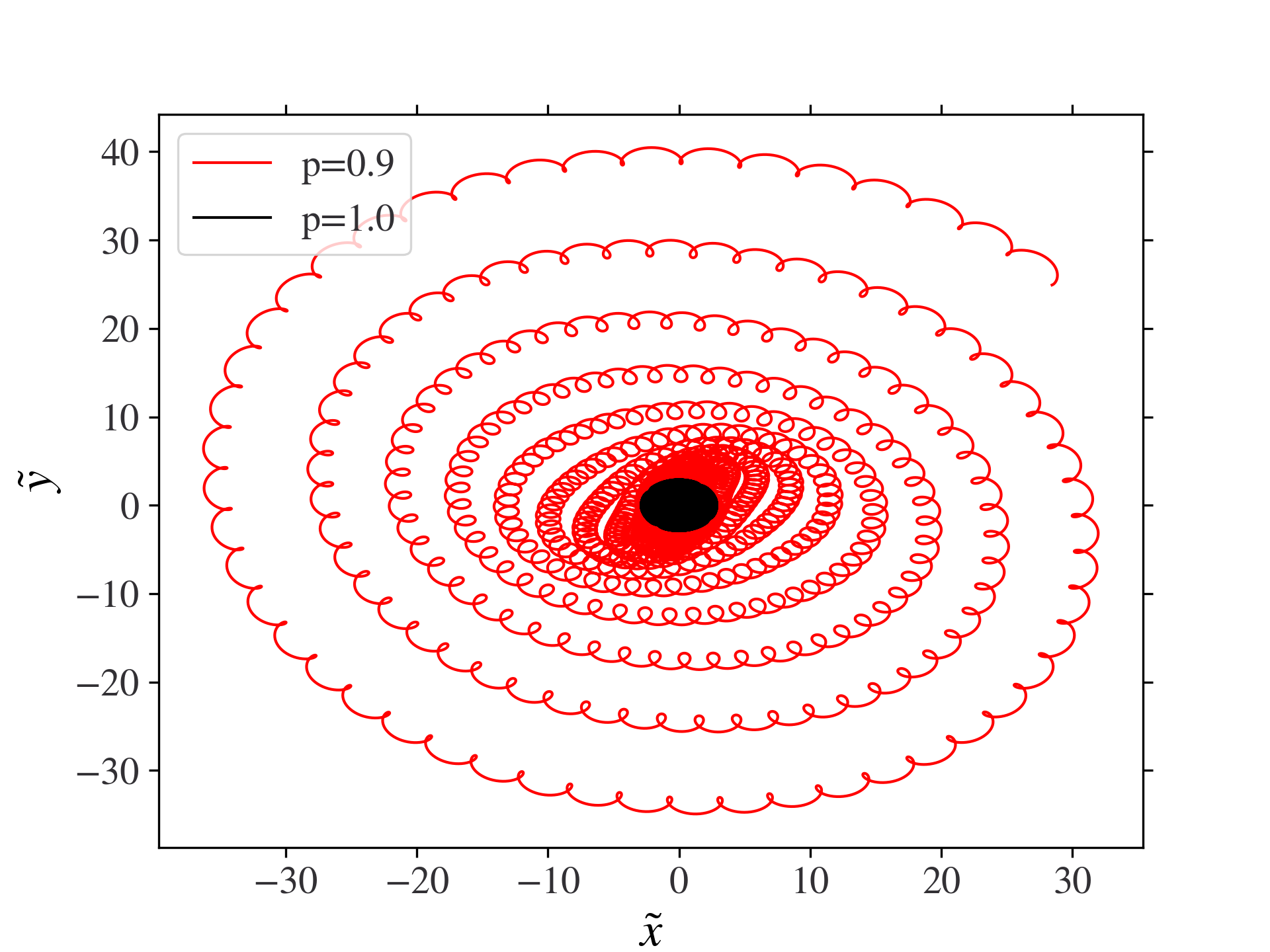}

}\subfloat[]{\includegraphics[scale=0.5]{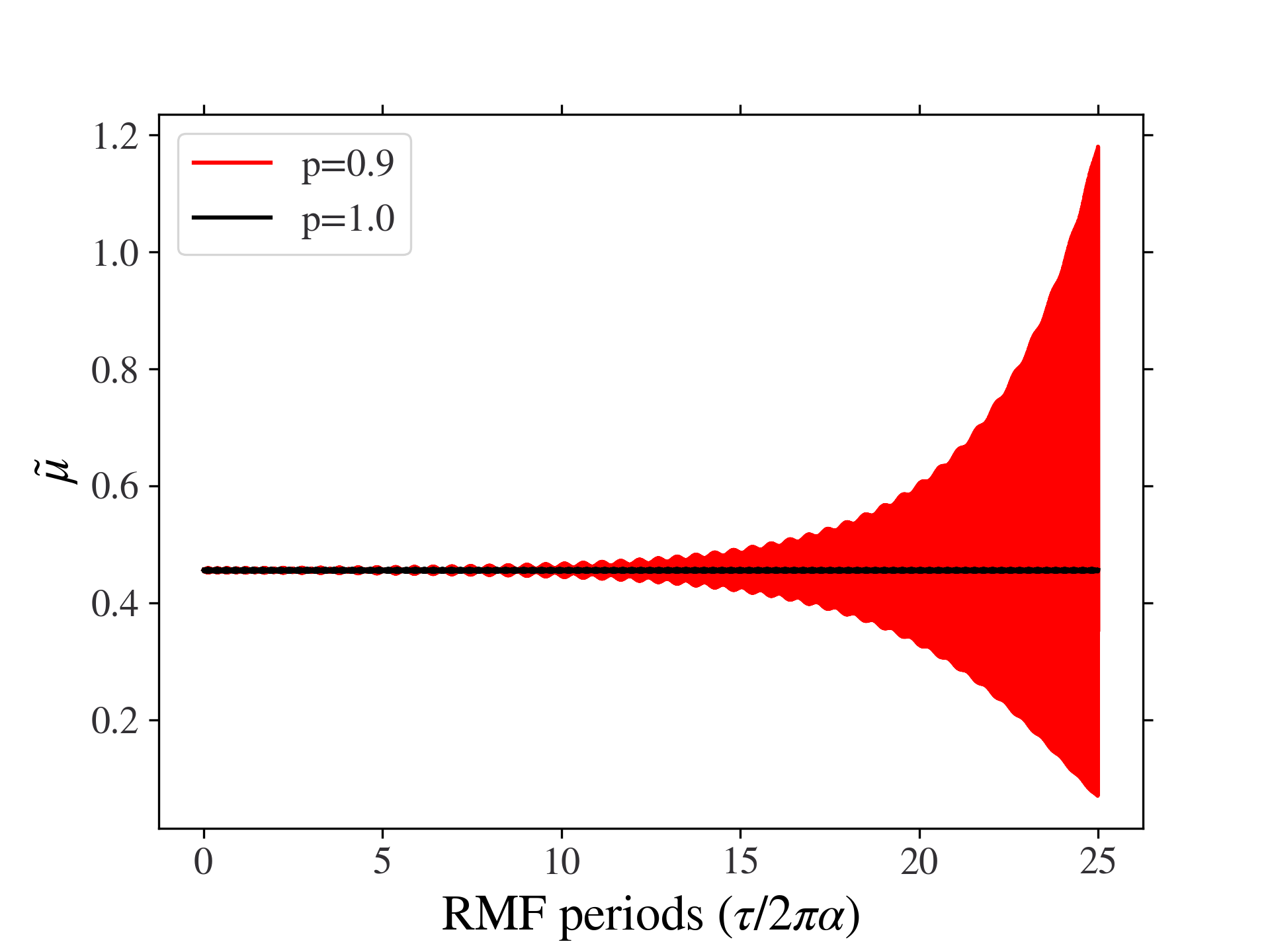}

}

\caption{(a) $\tilde{x}\tilde{y}$-projection of the trajectories and (b) magnetic
moments $\tilde{\mu}$ of test particle in a slow co-RMF with $\alpha=-50$,
$\beta=0.25$ over 25 RMF periods (1250 gyroperiods) for two different
boundary conditions: $p=1$ (red) and $p=0.9$ (black). The particle
is initialized at $\mathbf{\tilde{x}=0}$ and $\mathbf{\tilde{p}}=(1,0,0).$
These plots illustrate the structural instability of the p=1 dynamics
against perturbations in p. }
\label{fig:p1_trajectories} 
\end{figure}

The p=1 case is special and worth discussing since it has been previously
studied \citep{Soldatenkov1966,Fisch1982,Hugrass1983, Rax2016,Wetering2021}. In this case, $u_{1}=0$
is an exact root for all $(\alpha,\beta)$, as can be seen by the
vanishing of the constant term in the characteristic polynomial (\ref{eq:char_poly_u}).
In this case it may appear that the system is stable in the $\alpha\rightarrow-\infty$
limit since $u_{2}<0$. However, $u_{1}=0$ corresponds to a double
root at $\lambda=0$, and a closer inspection of the eigenvectors
shows the eigenspace of 0 is 1-dimensional and spanned by the vector
$\mathbf{e}_{z}$. Thus, 0 is a defective eigenvalue and the motion
is technically unbounded. This point was not reported in the literature.
This instability appears mild in the sense that the instability grows
linearly, rather than exponentially, in time. Morever, the instability
only cooresponds to growth of the $z$-coordinate; the motion in the
$xy$-plane and the momentum in all directions remain bounded as $\tau\rightarrow\infty.$
Both these properties are atypical, and only occur for the special
value of $p=1$; in all other cases we study, instabilities correspond
to the exponential divergence in both the axial and transverse coordinates.
This is in fact the fundamental issue with the $p=1$ case: small
modifications to the boundary conditions leading to nonzero electric
fields in the $xy$-plane destabilize the motion for large negative
$\alpha$ (see Fig.\,\ref{fig:p1_trajectories}). This is known as
structural instability. This sensitive dependence on the boundary
conditions needs to be considered when drawing conconclusions from
the $p=1$ case. It is shown in Sec.\,\ref{subsec:Geometric-Stability}
that the system is structurally stable for other $p$ values.

One potential complication in determining stability from asymptotic
expansions is that higher order terms could modify stability. The
only case in which this could be an issue is if all the $u$ are real
and negative to leading order with an imaginary term occurring at
some higher order. However, this is only possible when eigenvalues
are repeated in the leading order since complex eigenvalues occur
in conjugate pairs. For instance, higher order terms cannot effect
the stability in the above case when $p\neq0$ since the roots in
Eq.\,(\ref{eq:slow_sol}) are all distinct to leading order.

\subsection{Counter-rotating stability boundaries\label{subsec:Counter-rotating-boundaries}}

\begin{figure}
\includegraphics[scale=0.5]{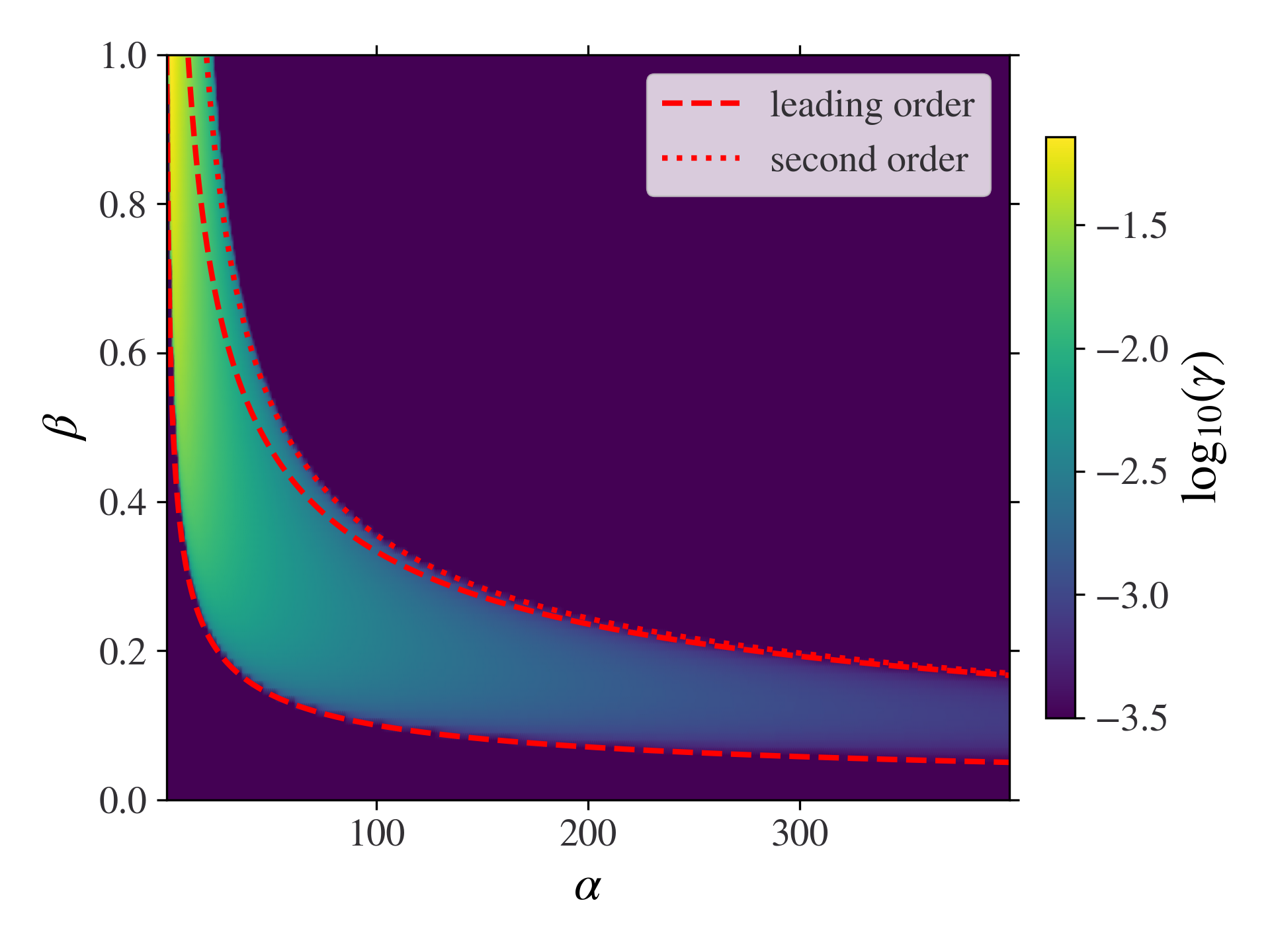}

\caption{Approximate stability boundaries for a co-rotating RMF for $p=-0.3$
are shown in red. A higher order approximation is needed for the upper
boundary when $\beta\sim O(1)$.}
\label{fig:counter_approx} 
\end{figure}

Figure \ref{fig:counter_stability} shows that in the counter-rotating
($\alpha>0)$ case, there are stability boundaries which depend on
$p$. By guessing the functional form $\beta\sim\frac{c}{\sqrt{\alpha}}$
of the boundary, and taking $\alpha\rightarrow+\infty$, one finds
\begin{align}
u_{1} & \sim-1+O(\alpha^{-1})\thinspace,\\
u_{2,3} & \sim\frac{1}{2\alpha^{2}}[pc^{2}-1\pm\text{\ensuremath{\sqrt{(c^{2}-1)(p^{2}c^{2}-1)}]}}+O(\alpha^{-3})\thinspace.
\end{align}
When $p<0$, stability boundaries correspond to changing signs of
the radical and are given to leading order by $c=1,\text{\ensuremath{\frac{1}{|p|}}}$,
that is, the stability boundaries are $\beta\sim\frac{1}{\sqrt{\alpha}},\frac{1}{|p|\sqrt{\alpha}}.$
When $0<p<1$, the dynamics again become unstable as $c$ increases
past 1 due to the radical. However, when the radical becomes real
again upon crossing $c=1/p$, the nonradical term $pc^{2}-1$ is positive,
maintaining the instability. Thus, the only stability boundary is
$\beta\sim\frac{1}{\sqrt{\alpha}}$. Analysis is similar when $p>1$,
except now the only stability boundary is $\beta\sim\frac{1}{p\sqrt{\alpha}}.$
We can summarize these results by saying that for $\alpha\gg1$ the
dynamics are unstable if 
\begin{align*}
\end{align*}
\begin{align}
\frac{1}{|p|\sqrt{\alpha}}<\beta<\frac{1}{\sqrt{\alpha}} & \text{ if }p<-1\thinspace,\\
\frac{1}{\sqrt{\alpha}}<\beta<\frac{1}{|p|\sqrt{\alpha}} & \text{ if }-1<p<0\thinspace,\\
\beta>\frac{1}{\sqrt{\alpha}} & \text{ if }0<p<1\thinspace,\\
\beta>\frac{1}{p\sqrt{\alpha}} & \text{ if }p>1\thinspace.
\end{align}

Using perturbation theory, one can obtain successively higher order
approximations of the stability boundaries. Doing so, one finds 
\begin{align}
\beta_{1} & \sim\frac{1}{\sqrt{\alpha}}+\frac{3-p}{4\alpha^{3/2}}\alpha^{-3/2}+O(\alpha^{-5/2})\thinspace,\\
\beta_{2} & \sim\frac{1}{|p|\sqrt{\alpha}}+\frac{(p-2)(p+1)}{4|p|^{3}\alpha^{3/2}}+O(\alpha^{-5/2})\thinspace.
\end{align}
The leading order approximation for $\beta_{1}$ is very accurate
even when $\alpha\sim O(1)$. For $\beta_{2}$, one needs to include
$O(\alpha^{-3/2})$ corrections when $\alpha\sim O(1)$. This is illustrated
in Fig.\,\ref{fig:counter_approx} for $p=-0.3$.

We note in particular that in the case of $p\approx-1$ when the electric
field is primarily in the $xy$-plane, the region of instability in
the counter-rotating case is extremely small, and thus bounded motion
for all time is possible for nearly all $\alpha,\beta$.

\subsection{Co-rotating stability boundaries\label{subsec:Co-rotating-boundaries}}

\begin{figure}
\includegraphics[scale=0.5]{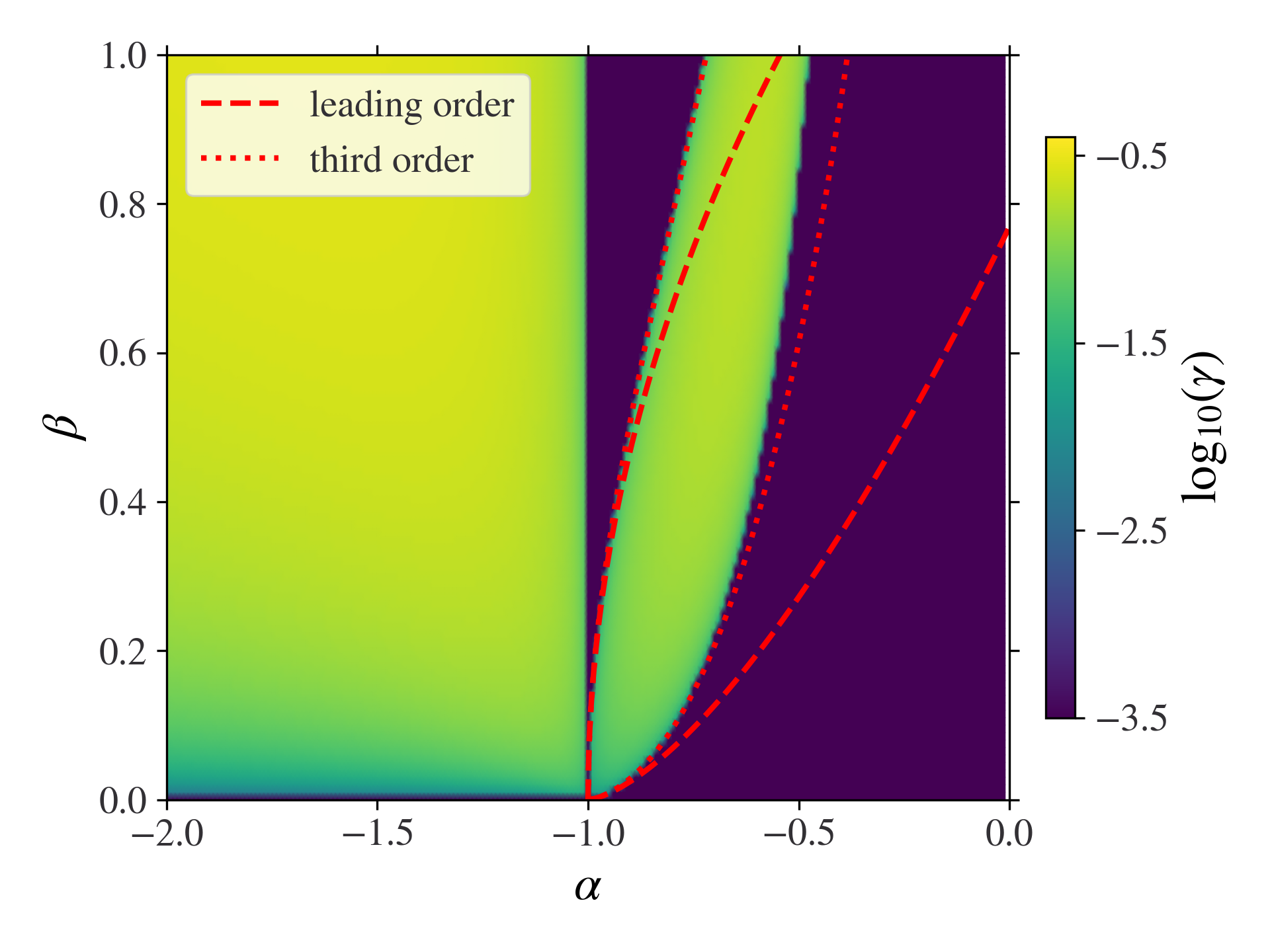}

\caption{Approximate stability boundaries in the co-rotating case for $p=-0.5$
are shown in red. Higher order approximations are necessary when $\beta\sim O(1)$.}
\label{fig:co_approx} 
\end{figure}

In the co-rotating case, the only stability boundaries are in the
fast RMF range $(-1\leq\alpha<0)$. Firstly, we show there is an exact
stability boundary at $\alpha=-1$ if we assume $-1\leq p\leq1$ and
$0<\beta<1$. If we let $\alpha=-1+\delta$, then the roots of the
characteristic polynomial are given to leading order in $\delta$
by 
\begin{align}
u_{1} & \sim\frac{p-1}{p+3}\delta\thinspace,\\
u_{2,3} & \sim-\frac{1}{2}\big(1+\beta^{2}\pm\sqrt{1+\beta^{2}(p^{2}+2p-1)+\beta^{4}}\big)\thinspace.
\end{align}
One can show that $u_{2,3}$ are always negative in our chosen parameter
regime, so the expression for $u_{1}$ implies that the system is
stable for $\delta>0$ and unstable for $\delta<0$. The stability
boundary at $\alpha=-1$ does not depend on the value of p, but there
are other stability boundaries which do. We can use asymptotic techniques
as in the previous section to obtain approximate equations for these
boundaries in the limit that $\beta^{2}\ll1$ and $|1+\alpha|\ll1$:
\begin{align}
\beta_{1} & \sim\frac{4\sqrt{1+\alpha}}{p+3}+\frac{9p^{2}+26p+29}{(p+3)^{3}}(1+\alpha)^{3/2}+O\big((1+\alpha)^{5/2}\big)\thinspace,\\
\beta_{2} & \sim\frac{(1+\alpha)^{3/2}}{1-p}+\frac{7-3p}{4(1-p)^{2}}(1+\alpha)^{5/2}+O\big((1+\alpha)^{7/2}\big)\thinspace.
\end{align}
These approximations are not very accurate when $\beta^{2}\sim1$,
but this can be partially remedied by including higher order corrections;
this is illustrated in Fig.\,\ref{fig:co_approx}.

\subsection{Extremely weak RMF limit}

\label{sec:weak_RMF} In the case that $\beta^{2}\ll\alpha^{-1}$,
the leading order solutions of Eq.\,(\ref{eq:char_poly_u}) are given
by 
\begin{align}
u_{1} & \sim\frac{(1-p)^{2}\beta^{2}}{4(1+\alpha)}+O(\beta^{4})\nonumber \\
u_{2} & \sim-\frac{1}{\alpha^{2}}+O(\beta^{2})\label{eq:Weak_RMF}\\
u_{3} & \sim-\frac{(1-\alpha)^{2}}{\alpha^{2}}+O(\beta^{2})\nonumber 
\end{align}
from which it follows that the dynamics are unstable only when $\alpha<-1$.

\section{Implications\label{sec:Implications}}

\subsection{p=0 case}

\begin{figure}
\subfloat[]{\includegraphics[scale=0.6]{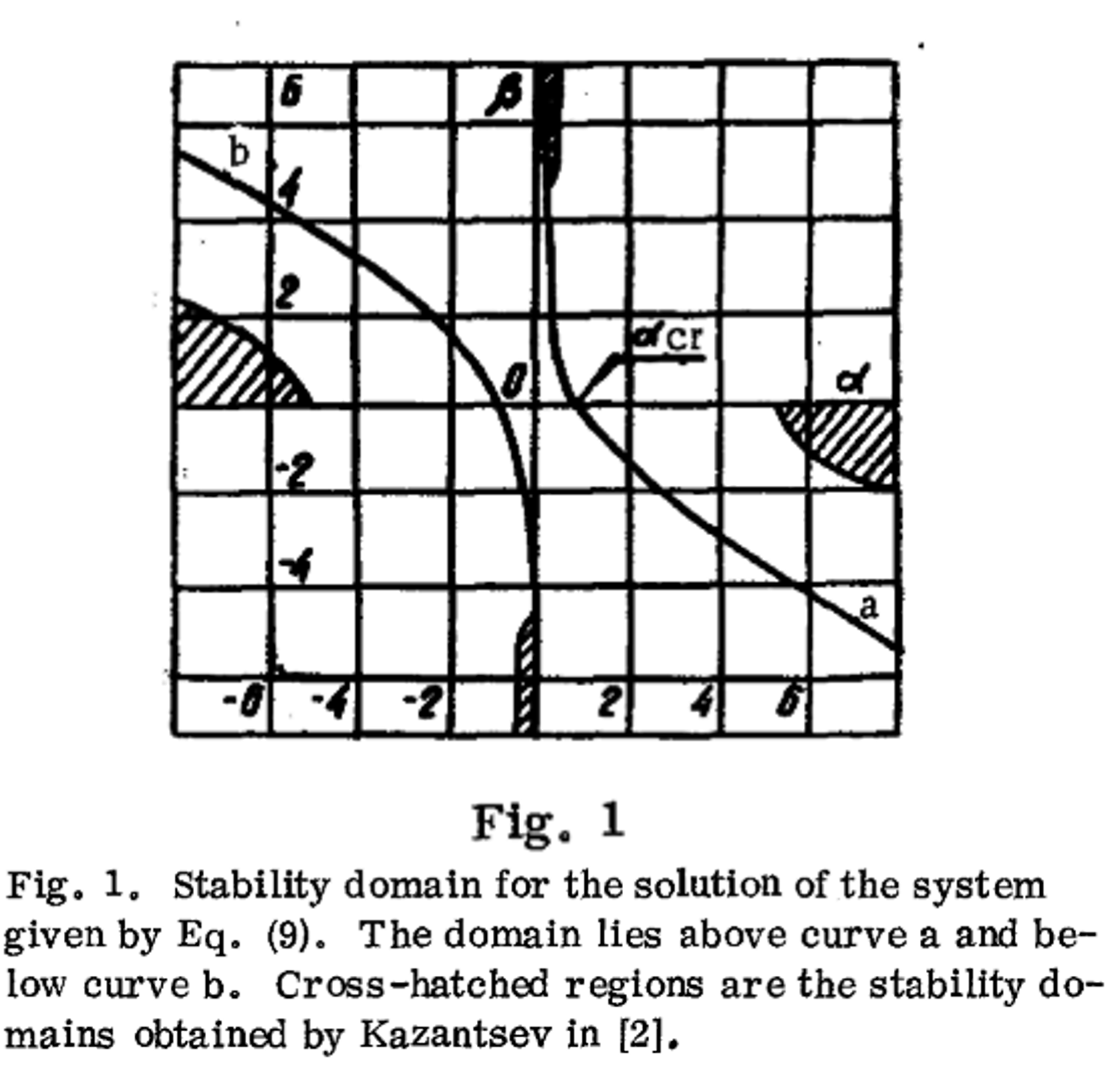}

}\subfloat[\label{p0correct}]{\includegraphics[scale=0.5]{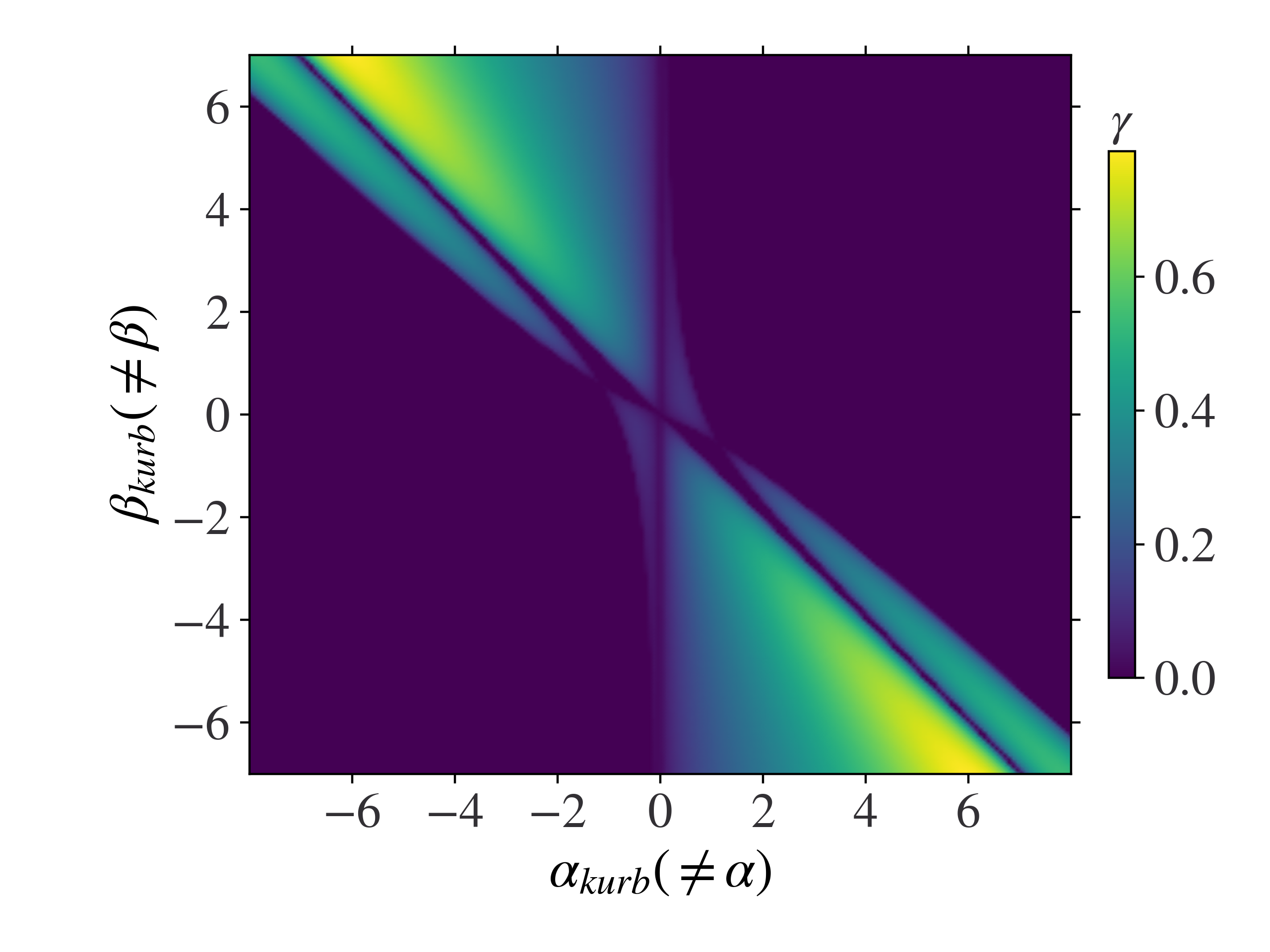}

}\caption{(a) Stability diagram obtained by Kurbatov parameterized
by $\alpha_{kurb}=\frac{1}{\alpha\beta}$ and $\beta_{kurb}=\text{\ensuremath{\frac{1}{\beta}}}$ (reproduced from Ref. \citep{Kurbatov1976} with permission of Springer Nature),
and (b) the corrected stability diagram calculated from Eq.\,(\ref{eq:char_poly_lam}).} 

\label{fig:kurbatov} 
\end{figure}

It appears that only the $p=0$ and $p=\pm 1$ cases have been treated
in the literature \citep{Kazantsev1959,Soldatenkov1966, Kurbatov1976,Fisch1982,Hugrass1982,Rax2016,Wetering2021}. The $p=0$ case corresponds to the electric field
\begin{equation}
\mathbf{E}=\frac{B_{1}\Omega_{1}}{2c}\big[-z\cos\epsilon\tau\mathbf{e}_{x}-z\sin\epsilon\tau\mathbf{e}_{y}+(x\cos\epsilon\tau+y\sin\epsilon\tau)\mathbf{e}_{z}\big]\thinspace
\end{equation}
and can be produced by a pair of dephased coils aligned along the x- and y-axes and a constant $\mathbf{B}_0$ along the z-axis. We remark on this case because its treatment in the literature is somewhat muddled. It was first considered by Kazantsev \citep{Kazantsev1959}, but an
an algebraic error in Eq. (3), Ref. \citep{Kazantsev1959} resulted in incorrect stability criteria. Kurbatov detailed this error
\citep{Kurbatov1976}, obtained the correct characteristic polynomial (Eq. 10, Ref. \citep{Kurbatov1976}), and produced a plot of the stability regions (Fig. 1, Ref. \citep{Kurbatov1976}). However, the boundaries in this plot  were distorted and certain stability boundaries were missing. Kurbatov's original plot and a corrected version produced from Eq. (\ref{eq:char_poly_lam}) are shown
in Fig.\,\ref{fig:kurbatov}; note that the parameters $\alpha_{kurb}$
and $\beta_{kurb}$ in Ref. \citep{Kurbatov1976} (subscripts added) are related to our definitions by $\alpha_{kurb}=\frac{1}{\alpha\beta}$ and $\beta_{kurb}=\frac{1}{\beta}$. A correct version of the $p=0$ stability diagram, consistent with Fig. \ref{p0correct}, can also be found in Fig. 4 of \citet{Rax2016}.

\subsection{Single-particle confinement in FRCs}

Previous analyses \citep{Fisch1982,Wetering2021} have used the
$p=1$ case to model single-particle confinement in field reversed
configurations (FRCs) maintained by RMFs. The idea in these analyses
is that RMF parameters $\alpha$ and $\beta$ should be chosen such
that both electrons and ions are confined for all time. Such conditions
would lead to long confinement times, aiding steady operation of an
FRC device. In these studies, it was assumed that $p=1$ and that
$|\alpha_{i}|\ll1$ and $|\alpha_{e}|\gg1$ where $\alpha_{e,i}=\text{\ensuremath{\frac{\Omega_{0e,i}}{\omega}}}$.
Van De Wetering and Fisch \citep{Wetering2021} propose that stable
operation can be achieved by applying an RMF which is counter-rotating
with respect to ions ($0<\alpha_{i}\ll1)$ and co-rotating with respect
to electrons ($|\alpha_{e}|\gg1,\alpha_{e}<0)$. There are a few issues
with this setup. For one, as was mentioned in Sec.\,\ref{subsec:Slowly-rotating-limit},
the $z$ coordinate actually grows linearly when $p=1$, and so the
motion is not technically bounded. This would not likely be a practical
issue since mirror forces could confine electrons in the $z$-direction.
The structural instability is a more important issue--if $p$ is
perturbed even slightly, the motion is unstable in all position and
momentum coordinates if $|\alpha_{e}|\gg1,\alpha_{e}<0$. Furthermore,
it appears that the reason $p=1$ was chosen is that it leads to nice
analytically solvable equations. Physically, it corresponds to the
situation in which $\mathbf{E}\parallel\hat{\mathbf{e}}_{z}$. However,
a model of RMFs in FRCs using more realistic nonlinear fields involves
azimuthal electric fields \citep{Glasser2002}.

We thus look for other parameter regimes which may result in confinement.
For all $p\neq1$, electrons are unstable when $|\alpha_{e}|\gg1,\alpha_{e}<0$
due to what we might term the co-rotating electromagnetic instability.
It is only possible to confine electrons for all time in the $|\alpha_{e}|\gg1$
limit if the RMF is in fact counter-rotating with respect to the electrons
($\alpha_{e}>0)$. In this case, either of the following conditions
would lead to stable electron motion: 
\begin{align}
\beta & >\frac{\max(1,|p|^{-1})}{\sqrt{\alpha_{e}}}\text{ and }p<0\thinspace\\
\text{or}\nonumber \\
\beta & <\frac{\min(1,|p|^{-1})}{\sqrt{\alpha_{e}}}\thinspace.
\end{align}
Assuming $|\alpha_{i}|\ll1$ and $\alpha_{i}<0$, the ions are also
stable.

We note that such considerations should only be applied to the edge
of the FRC where the magnetic field gradients are relatively weak
and the approximation of a spatially uniform magnetic field is valid.
This model is not applicable in the core of the FRC due to the presence
of large field gradients.

\subsection{Heating particles with RMFs}

Because we are working in a linear model, growth in the position and
momentum coordinates are coupled. In particular, it is not possible
for particle energy to grow while maintaining a bounded orbit. This
can be seen by noting that $\mathbf{M}'$ in Eq.\,(\ref{eq:M_prime})
cannot have eigenvectors containing only the momentum coordinates.
The studies \citep{Fisch1982,Wetering2021} only considered situations
in which all particle dynamics are stable, and therefore, particle
energies are bounded. In this case, particle energies oscillate over
an RMF period, and thus heating can occur if one considers phase de-cohering
collisions, whether with particles or the field. In fact, this is
the only way heating can occur when particle orbits are bounded in
this model.

However, it seems that this requirement on particle orbits may be
too restrictive. The fact that bounded orbits cannot exist when heating
occurs is a result of the simplified, linear model employed. It is
a characteristic of any linear instability, including ICRF heating
which is frequently used to heat laboratory plasmas. In realistic
scenarios, nonlinear fields and collisions could nonlinearly saturate
the instability and provide confinement. FRCs, for example, have axial
field nulls, around which the model we are considering is certainly
not applicable. Furthermore, ambipolar electric fields are likely
to develop near the boundary to limit particle losses. Thus, we can
instead look for parameter regimes in which electrons and/or ions
have unstable orbits and thus undergo collisionless heating, and assume
that confinement is provided by nonlinear effects outside of our model.
There are broad parameter regimes in which such heating can occur.
For example, heating for both species would occur if the RMF is taken
to be co-rotating with ions and counter-rotating with electrons, and
such that $\alpha_{i}\gtrsim1$ and $\alpha_{e}>\beta^{-2}$ if we
assume $p>0$.

We can give a physical picture of the energization mechanism. Since
magnetic fields can do no work, it is the inductive electric field
which imparts energy to the particles. In the case of $p=1$, the
particle is accelerated in the $z$-direction by the purely axial
electric field. The radial RMF then helps convert this axial motion
into perpendicular motion, increasing $\mu$. The details are different
for $p\neq1$, but the same essential idea that the electric field
imparts energy and the RMF redistributes it among the other degrees
of freedom holds.

As in the previous section, this model should only be applied to the
edge of an FRC plasma, where the field gradients are relatively weak.
In particular, the field-nulls inside the separatrix appear to play
an important role in heating in the core of the FRC \citep{Glasser2002},
and thus such heating cannot be modeled by the spatially uniform magnetic
field employed here. We can apply this theory to the edge of the PFRC-2
near the separatrix, where typical parameters are $\alpha_{e}=-1.8\times10^{3}$,
$\alpha_{i}=0.9$, and $\beta=0.8\times10^{-3}$. The model predicts
electron heating in the edge since the electrons are co-rotating with
the RMF and $|\alpha_{e}|\gg1$ (see figure \ref{fig:co_stability}).
Since $\alpha_{i}\ll\beta^{-2}$, the ions are stable by the results
of section \ref{sec:weak_RMF}, and thus ion heating due to this mechanism
is not expected in the edge.

\subsection{Azimuthal current drive}

The rotamak concept \citep{Jones1990,HUgrass1980} relies on RMFs
to provide the large azimuthal current necessary to maintain field-reversal
in an FRC. The heuristic picture of the RMF current drive mechanism
is that electrons are tied to the RMF and are dragged azimuthally
by it. The ions are too heavy and slow to be magnetized by the RMF,
resulting in a net azimuthal current due to the electrons. Studies
of single-particle motion in non-linear FRC-like fields suggest this
picture is an oversimplification of the current drive mechanism \citep{Glasser2001b}.
However, our model suggests that with the help of the instability
this mechanism may be accurate during the initial formation of an
FRC.

When the slow RMF is first applied, prior to FRC formation, the electromagnetic
fields are approximated by Eqs. (\ref{eq:RMF_field}) and (\ref{eq:E_field}).
Assume $\alpha_{e,i}$, $\beta$, and $p$ are such that electrons
are unstable, ions are stable, and the RMF is co-rotating with respect
to the electrons. Let $\mathbf{\xi}$ be the position-space projection
of the eigenvector of $\mathbf{M}'$ corresponding to the most unstable
eigenvalue $\lambda_{u}$ for electrons. Eq. (\ref{eq:slow_sol})
shows that $\lambda_{u}$ is real to leading order as $|\alpha|\rightarrow\infty$,
and thus $\gamma\approx\lambda_{u}.$ This eigenmode will eventually
dominate the electron dynamics with the asymptotic motion in the rotating
reference frame given by $\mathbf{x}'(\tau)\sim a\mathbf{\xi}e^{\gamma\tau}$
where $a$ is a constant determined by initial conditions. In the
lab-frame, we see that all electrons rotate synchronously with the
RMF and with an exponentially growing radius, 
\begin{equation}
\tilde{\mathbf{x}}(\tau)\sim a\mathbf{R}(-\epsilon\tau)\mathbf{\xi}e^{\gamma\tau}.\label{eq:current_drive}
\end{equation}
Consider a ring of such electrons with initial radius $r_{0}$ and
extended uniformly in the $z$-direction. Even if the particles are
initially stationary, Eq. (\ref{eq:current_drive}) shows that the
particles will eventually establish an azimuthal current, creating
a solenoid with radius $r=r_{0}e^{\gamma\tau}$ and current $K$ per
unit length. Since the asymptotic rotation speed is synchronous with
the RMF, $K$ will be constant while the radius grows. Thus, the linear
instability in the electrons results in a growing enclosed magnetic
flux, 
\begin{equation}
\Phi=-\frac{4\pi^{2}}{c}Kr_{0}^{2}e^{2\gamma\tau}.
\end{equation}
The ions, being bounded, will produce no such growing magnetic flux,
resulting in net azimuthal current drive. We observe that the instability
significantly enhances the azimuthal current drive effect of an RMF
through two mechanisms. The instability synchronizes the angular velocity
of all electrons to that of the RMF, and pushes all electrons toward
larger radii.

\subsection{Geometric stability\label{subsec:Geometric-Stability}}

In this analysis we have assumed that the boundary conditions rotate
with the the same phase and frequency as the RMF. This assumption
allowed us to parameterize the boundary conditions by a single parameter
$p$. Furthermore, these assumptions were necessary to obtain a time-independent
set of equations after transforming to the rotating frame. These assumptions
cover many situations of interest, such as the RMFs typically applied
to cylindrical devices like FRCs and mirror machines. There are other
situations in which it may be necessary to consider boundary conditions
which break these assumptions, for example, the rotational symmetry
would be broken in the case of RMFs applied to a toroidal device.
In this case, one may wish to work in the large-aspect ratio approximation
in which the rotationally symmetric boundary condition assumption
will only be perturbatively broken. Furthermore, even in situations
like FRCs where the boundary conditions should be rotationally symmetric,
one needs to be careful that the stability diagram itself is structurally
stable against small perturbations that break this symmetry.

When considering general boundary conditions, transforming to the
rotating reference frame no longer eliminates time-dependence in the
equations of motion. As such, the eigenvalues of the monodromy matrix
$\mathbf{P}(\tilde{T})$ determine stability, rather than those of
$\mathbf{M}(\tau)$ or $\mathbf{M}'(\tau)$. In particular, the system
is stable if the eigenvalues $\rho_{i}$ of $\mathbf{P}(\tilde{T})$
are in the unit disk, with those on the boundary semisimple. Since
$\mathbf{M}(\tau)$ is Hamiltonian (i.e. $\mathbf{M}(\tau)\in\mathfrak{sp}(2n,\mathbb{R}))$,
$\mathbf{P}(\tilde{T})\in\text{Sp(2n,\ensuremath{\mathbb{R})}}$.
As a result, the Krein theory applies---stable eigenvalues must reside
on the unit circle, and can only leave and become unstable via a Krein
collision \citep{Krein1950,Gelfand1955,Moser1958}. Consider a set
of boundary conditions that satisfy our assumptions (i.e. they can
be specified by choosing $p$) and which are stable for some choice
of $\alpha$ and $\beta$. The eigenvalues $\rho_{i}$ are on the
unit disk and semisimple. The Krein theory thus guarantees that arbitrarily
small perturbations to the boundary conditions, including those which
cannot be parameterized by $p$, cannot destabilize the system. Note,
however, that in the marginal case when there are non-semisimple eigenvalues
on the unit circle, perturbations can cause those eigenvalues to leave
the unit circle, making the instability worse. This is what occurs
in the $p=1$ case. However, for generic p values, we can say that
the regions of stability are geometrically protected, since the Hamiltonian
(or symplectic) structure provides this structural stability.

\section{Instability in a parallel oscillating field\label{sec:Stability-in-a}}

\begin{figure}
\subfloat[\label{mathieu_stability}]{\includegraphics[scale=0.5]{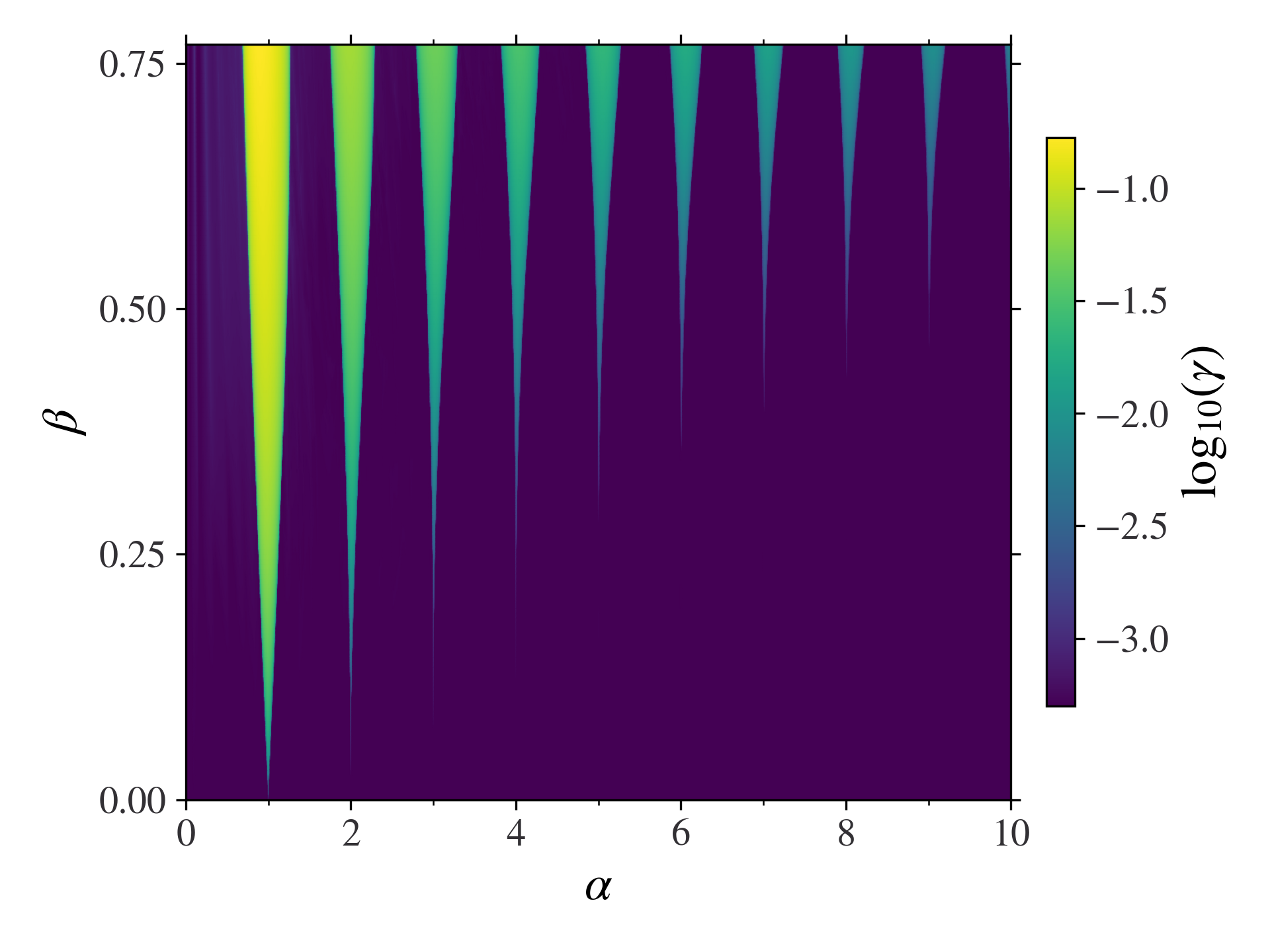}

}\subfloat[\label{mathieu_AI}]{\includegraphics[scale=0.5]{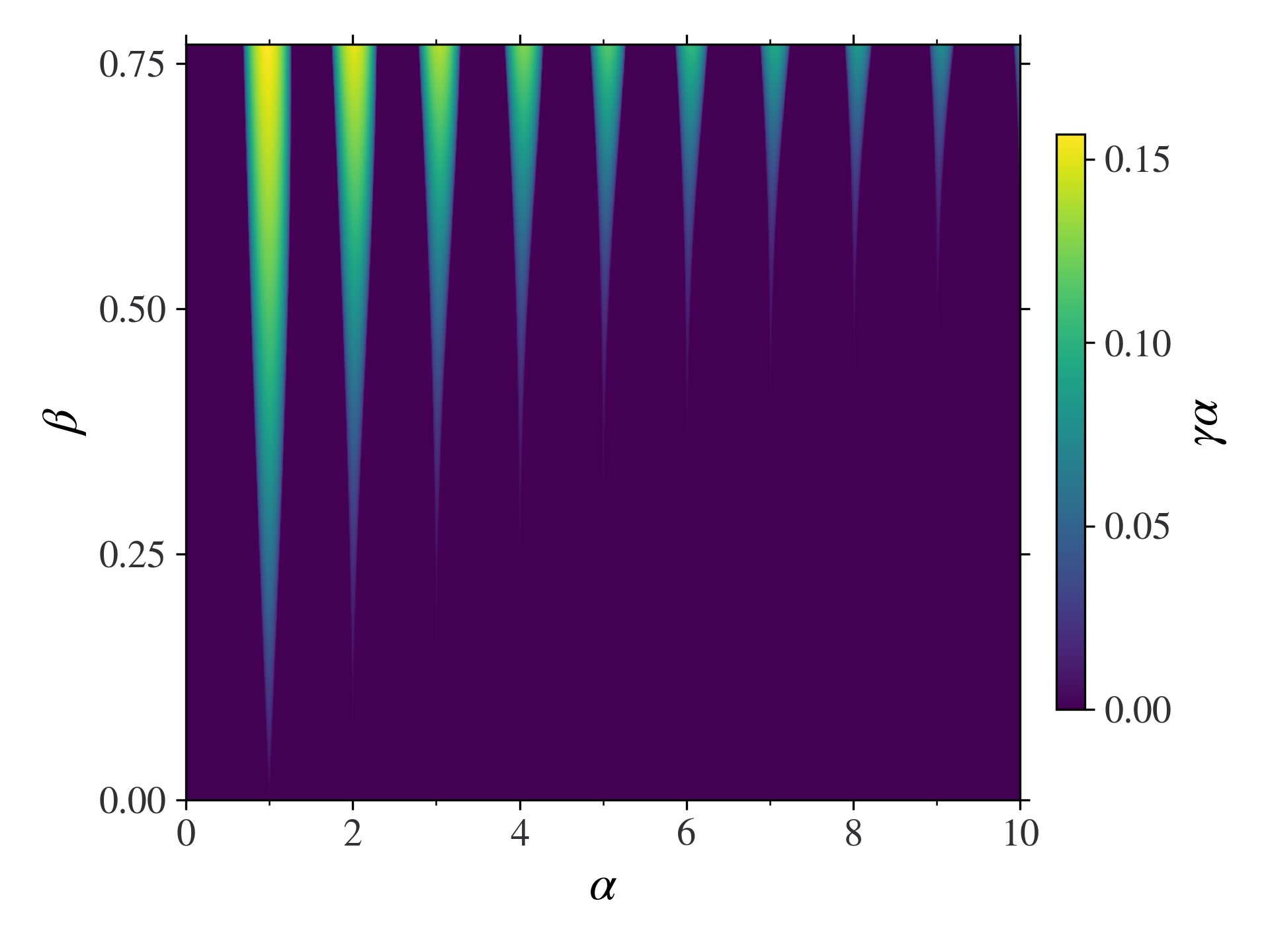}

}

\caption{(a) Stability diagram of a particle in a parallel oscillating magnetic
field parameterized by the normalized RMF strength $\beta$ and inverse
frequency $\alpha$. The unstable band structure is related to that
of the Mathieu equation. (b) Plot of the one-RMF-period growth rate
$\gamma\alpha$ as a function of $\alpha$ and $\beta$. The decay
of $\gamma\alpha$ as $\alpha$ increases indicates adiabatic invariance.}
\label{fig:alpha_gamma_parallel} 
\end{figure}

The instabilities in $\mu$ and corresponding energization of particles
observed in the slow RMF fields of Eq.\,(\ref{eq:RMF_field}) are
surprising in light of the adiabatic invariance of the magnetic moment.
To facilitate the discussion of adiabatic invariance in the next section,
we consider an additional, simpler magnetic configuration, that in
which the periodic magnetic field $\mathbf{\tilde{B}}=\tilde{B}(\tau)\mathbf{e}_{z}$
is purely in the $z$-direction so that the motion is confined to
the $xy$-plane \citep{Qin06PRL}. The normalized equations of motion
are 
\begin{align}
 & \ddot{\tilde{x}}=\frac{1}{2}\tilde{y}\dot{\tilde{B}}+\dot{\tilde{y}}\tilde{B}\thinspace,\\
 & \ddot{\tilde{y}}=-\frac{1}{2}\tilde{x}\dot{\tilde B}-\dot{\tilde{x}}\tilde{B}\thinspace.
\end{align}
Unlike the RMF case, there is no obvious way to eliminate the time-dependence
in the EOMs. However, they can be reduced to a 1-dimensional Hill
equation as done by Ogawa \citep{Ogawa1962}. Letting $\xi=\tilde{x}+i\tilde{y}$,
one obtains the complex scalar equation 
\begin{equation}
\ddot{\xi}+i\tilde{B}\dot{\xi}+\frac{i}{2}\dot{\tilde{B}}\xi=0.
\end{equation}
Defining $\xi(\tau)=u(\tau)\exp\Big[-\frac{i}{2}\int_{0}^{\tau}\tilde{B}(t)dt\Big]$
puts this into the form of Hill's equation, 
\begin{equation}
\ddot{u}+\frac{1}{4}\Omega^{2}(\tau)u=0.\label{eq:Hill1}
\end{equation}
Since $|\xi|=|u|$, $\xi$ and $u$ have the same stability properties.
Choosing $\Omega(\tau)=1+\beta\sin\epsilon\tau$, we have 
\begin{equation}
\ddot{u}+\frac{1}{4}\Big(1+\beta\sin\epsilon\tau\Big)^{2}u=0.\label{eq:Hill2}
\end{equation}
This equation can be viewed as a generalization of Mathieu's equation
with the proper Mathieu's equation recovered in the small $\beta$
limit. Such equations are difficult to study analytically, in part
because they are non-hypergeometric \citep{Higham2015}. We instead
calculate the instability rates numerically from the monodromy matrix
$\mathbf{P}(\tilde{T})$ using methods that have been employed to
study other non-autonomous instabilities in plasma \citep{Qin2014TwoStream,Qin2014}
and accelerator physics \citep{Davidson01-all,Qin2010PRL,Qin2013PRL2,Qin2019LH}.
The instability rate $\gamma$ is determined by the eigenvalues $\rho_{i}$
of $\mathbf{P}(\tilde{T})$: 
\begin{equation}
\gamma\mathrel{\dot{=}}\frac{\ln(\max|\rho_{i}|)}{\tilde{T}}.\label{gamma_rho}
\end{equation}
The stability diagram in Fig.\,\ref{mathieu_stability} shows that
the unstable regions form a band structure of so-called Arnold tongues
centered around the cyclotron resonances at integer values of $\alpha=\epsilon^{-1}$.
Note that all Arnold tongues are connected to the line of $\beta=0$,
i.e., instability regions exist around arbitrarily large integer values
of $\alpha$ and arbitrarily small $\beta$, akin to the situation
of Mathieu's equation. However, for fixed $\beta$, these unstable
regions shrink rapidly in size and strength as $\alpha$ increases,
and thus such fields cannot easily energize particles for large $\alpha$.
In comparison, it is remarkable that instability in the slowly RMF
discussed in the previous section exists for large regions in the
parameter space. The existence of such parametric resonances for arbitrarily
slow driving forces is prototypical of the subtle issues that arise
in the theory of adiabatic invariants \citep{Chirikov1978, Arnold1989}.

\section{Adiabatic Invariance\label{sec:Adiabatic-Invariance}}

\begin{figure}
\subfloat[\label{alpha_gamma_co}]{\includegraphics[scale=0.5]{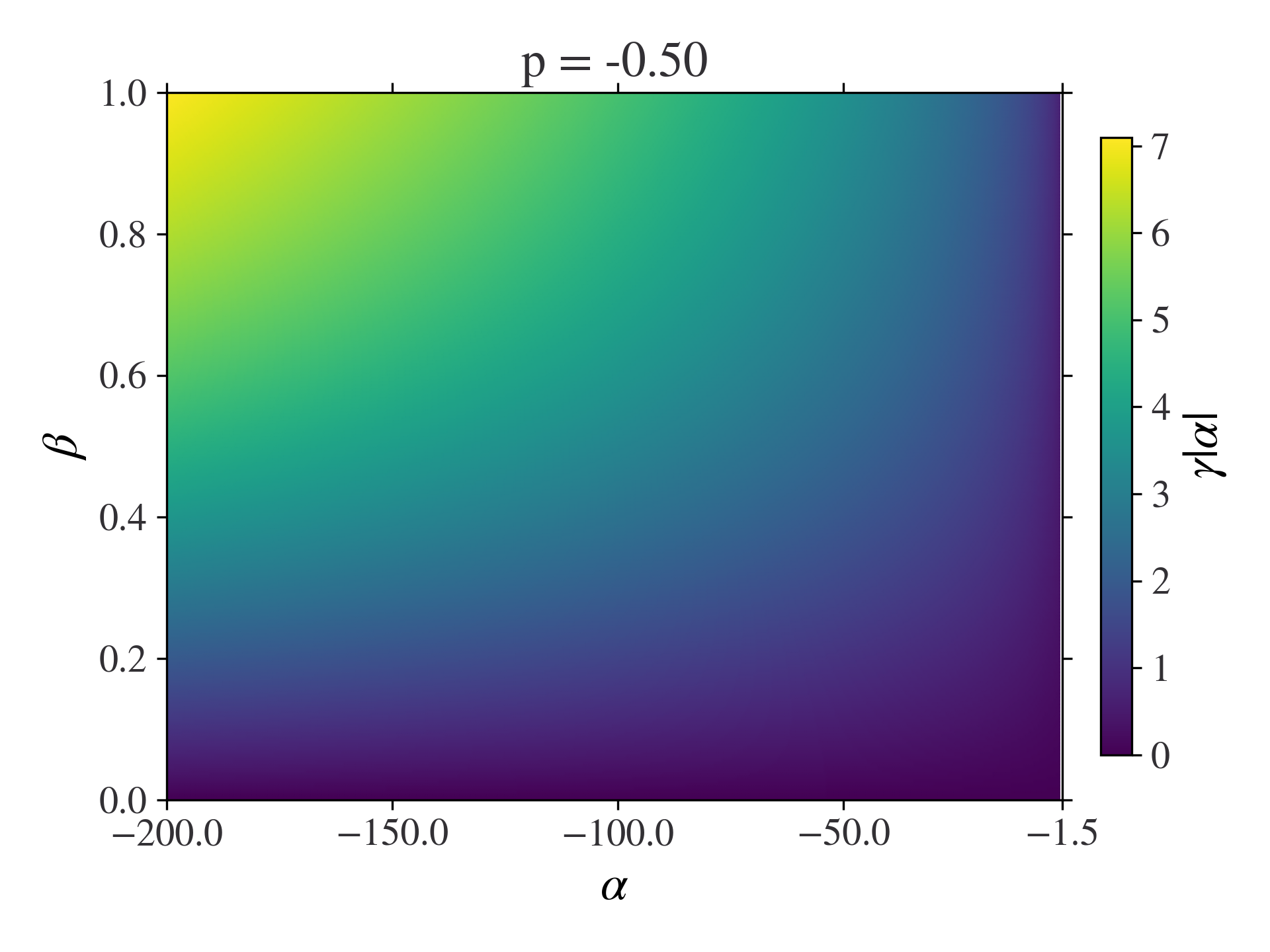}

}\subfloat[\label{alpha_gamma_counter}]{\includegraphics[scale=0.5]{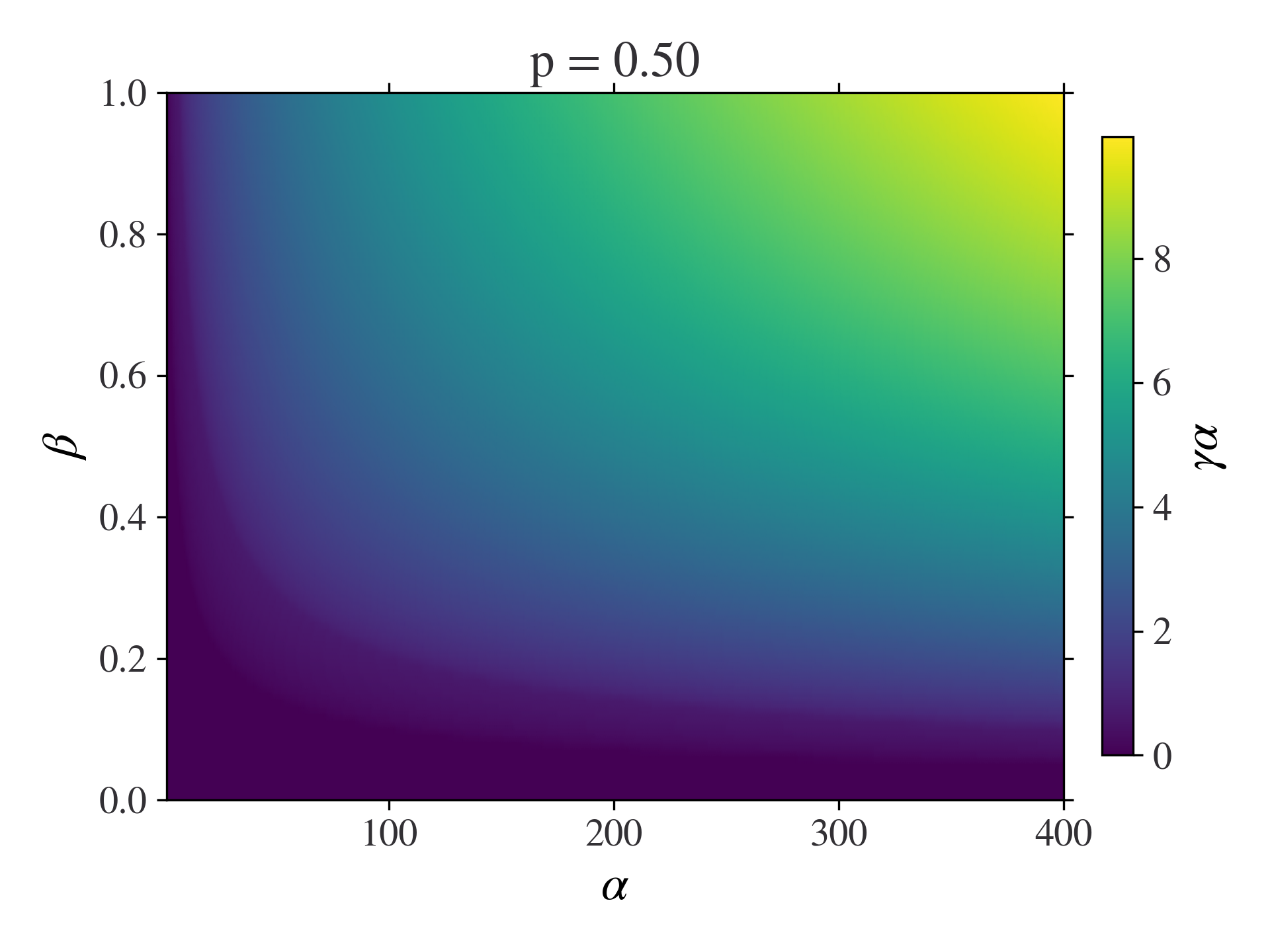}

}

\caption{Plots of the one-RMF-period growth rate $\gamma\alpha$ as a function
of RMF parameters $\alpha$ and $\beta$ for (a) a co-RMF with $p=-0.5$
and (b) a counter-RMF with $p=0.5$.}
\label{fig:alpha_gamma_RMF} 
\end{figure}

In this section, we address the seeming conflict between the adiabatic
invariance of $\mu$ and instabilities of $\mu$ in the slowly varying
RMF and parallel magnetic field discussed in previous sections. We
show how these results are in fact consistent. The adiabatic invariance
of the magnetic moment refers to the tendency of variations in $\mu$
to remain small over long time periods in the asymptotic limit of
slowly varying fields. For quantitative analysis, it is necessary
to give this concept a rigorous definition; one sufficient for the
current discussion is given by Arnold \citep{Arnold1989}. Let $H(\mathbf{q},\mathbf{p};\eta)$
be a fixed, $C^{2}$ function of $\eta$. Set $\eta=\epsilon\tau$
and consider the Hamiltonian evolution with slowly varying parameter
$\eta$: 
\begin{equation}
\mathbf{\dot{q}}=\frac{\partial H}{\partial\mathbf{p}},\;\;\mathbf{\dot{p}}=-\frac{\partial H}{\partial\mathbf{q}};\;\;H=H(\mathbf{q},\mathbf{p},\epsilon\tau).
\end{equation}
Then a function $I(\mathbf{q},\mathbf{p};\eta)$ is an adiabatic invariant
of this system if for every $\delta>0$ there exists $\epsilon_{0}>0$
such that if $0<\epsilon<\epsilon_{0}$ and $0<\tau<1/\epsilon$,
then 
\begin{equation}
|I(\mathbf{q}(\tau),\mathbf{p}(\tau);\epsilon\tau)-I(\mathbf{q}(0),\mathbf{p}(0);0)|<\delta.
\end{equation}
In the present analysis in which we are interested in the limit of
slowly varying fields ($\epsilon=\alpha^{-1}=\frac{\Omega_{1}}{\Omega_{0}}\ll1$),
adiabatic invariance of $\mu$ is equivalent to the claim that for
every $\delta>0$ there is a $\epsilon_{0}$ such that 
\begin{equation}
|\mu(\tau;\epsilon)-\mu(0;\epsilon)|<\delta
\end{equation}
for all $0<\epsilon<\epsilon_{0}$ and $\tau<\tilde{T}\sim|\epsilon|^{-1}=|\alpha|$.
For the linear fields studied, we have the following bound on the
growth of $\mu$ 
\begin{equation}
\mu(\tau)\sim\tilde{v}_{\perp}^{2}(\tau)\leq\tilde{v}^{2}(\tau)\leq\tilde{v}^{2}(0)e^{2\gamma\tau}\,,
\end{equation}
so adiabatic invariance is equivalent to the condition that for any
fixed $\beta$,

\begin{equation}
\lim_{|\alpha|\rightarrow\infty}\gamma(\alpha,\beta)\alpha=0.\label{eq:AI}
\end{equation}
Physically, this requirement is that the growth factor over a single
RMF period must approach $0$ as the RMF period becomes arbitrarily
large. The asymptotic nature of this condition is enough to resolve
any seeming contradiction between particle energization and adiabatic
invariance. The fundamental observation is that while this growth
rate must eventually approach $0$ as $|\alpha|\rightarrow\infty$,
the instability rate can be finite for fixed $\alpha$, and may even
increase over a range of $\alpha$ values. In the case of the parallel
magnetic field in the previous section, unstable regions exist for
arbitrarily large $|\alpha|$, but the magnitude of $\gamma\alpha$
decreases rapidly as $|\alpha|$ increases (see Fig.\,\ref{mathieu_stability}).
The situation is quite different for some types of RMFs. In the co-rotating
case with $p<0$, Eq.\,(\ref{eq:slow_sol}) shows that $\gamma|\alpha|$
approaches a constant as $|\alpha|\rightarrow\infty$. Even more dramatically,
in both the co- and counter-rotating cases with $p>0$ and $p\neq1$,
the same equations show that $\gamma\alpha\sim O\big(|\alpha|^{1/2}\big)$;
see Fig.\,\ref{alpha_gamma_counter}. Such apparent violations of
adiabatic invariance are an artifact of the linearity and infinite
extent of the fields. The classical results on the adiabatic invariance
of $\mu$ \citep{Kruskal1958b,Kruskal1962c,Berkowitz1959b} require
that $\mathbf{E}$ and $\mathbf{B}$ are bounded in $\mathbf{x}$,
which is violated by the linear field $\mathbf{E}=-\frac{1}{c}\dot{\mathcal{A}}\mathbf{x}$.
Physically, $\mathbf{E}$ must decay for large $\mathbf{x}$, leading
to nonlinear saturation of the instability, and ultimately restoring
adiabatic invariance in $\alpha$. In essence, adiabatic invariance
ensures that the non-decreasing trend of $\gamma\alpha$ in Fig. \ref{fig:alpha_gamma_RMF}
will not persist for arbitrarily large $|\alpha|$. The size of $|\alpha|$
where this eventual downturn occurs will depend on the realistic nonlinear
fields but can theoretically be large.

We emphasize, however, that the necessity of the boundedness condition
on $\mathbf{E}$ and $\mathbf{B}$ is not obvious. Indeed, no analogous
requirement is needed to prove the existence of adiabatic invariants
for 1D Hamiltonian systems \citep{Arnold1989}. For example, the energy
to frequency ratio in the 1D simple harmonic oscillator is the model
example of an adiabatic invariant, and that system has a linear, and
thus unbounded, force. In fact, the observed adiabatic invariance
of $\mu$ in the parallel oscillating magnetic field case was not
guaranteed by the classical results on $\mu$ invariance, but rather
by results on adiabaticity in 1D Hamiltonian systems since the dynamics
are described by the scalar Hamiltonian system (\ref{eq:Hill2}).
The RMF case demonstrates the necessity of the boundedness of the
fields in the fully 3D case. The constraints of adiabatic invariance
thus do not set in until nonlinear effects become important. However,
that is also the condition in which nonlinear saturation typically
occurs for any linear instability. This gives a further reason why
adiabaticity may be a milder constraint in some scenarios than is
often assumed. We note too that, as pointed out by Arnold \citep{Arnold1989},
results on adiabatic invariance typically rely on certain averaging
techniques which may not be rigorously justified in dimensions higher
than 1. Thus, it is possible that subtle mathematical issues beyond
the scope of this paper may be relevant in this discussion.

\section{Discussion and Conclusions \label{sec:Conclusion}}

We have shown that it is possible to pump energy into a magnetized
particle via a slowly rotating magnetic field, despite the adiabatic
invariance of $\mu$. In practice, adiabatic invariance is only important
in as much as the associated asymptotic limit is realized. How large
$|\alpha|$ needs to be before adiabaticity sets in depends very much
on specifics of the field configuration. This variance is illustrated
by the difference between the parallel and rotating magnetic field
cases, with the former showing narrow resonances that shrink and weaken
rapidly as $|\alpha|\rightarrow\infty$ but the latter exhibiting
broadband instability that decreases slowly in this limit. This broadband
instability illustrates how adiabatic invariance may be significantly
less restrictive in practice than it appears.

It was shown that the existence and location of these unstable regions
depends critically on the boundary conditions. Even restricting to
the experimentally relevant boundary conditions parameterized by $p$,
one can have fairly strong control over the stability diagram. By
appropriately choosing RMF parameters and boundary conditions it is
theoretically possible to energize one or both species in a plasma.
This offers a simple model that could describe electron and ion heating
in the edge of FRCs driven by RMFs. Furthermore, in certain parameter
regimes these instabilities can drive azimuthal current, a mechanism
which may be important in the formation of FRCs. Accurate predictions
would of course require modeling with more realistic nonlinear field
configurations; the linear case presented here suggests such efforts
may be fruitful. Moreover, there are many types of FRCs: large or
small; pulsed or steady state; RF, beam, or compression heated; metal
or dielectric containment vessels; and research-device or reactor
scale. As such there is a broad range of $\beta$ and $\alpha$ values.
FRCs also contain plasma, with its concomitant ambipolar constraint,
and have magnetic nulls and strong field gradients. All these must
be included in the evaluating the applicability of the above analyses
to a particular FRC device.

In the present analysis of particle dynamics, no collisions are included.
For the physics of azimuthal current drive by RMF, we plan to investigate
the effects of collisions by numerically integrating the stochastic
differential equation for electron pitch-angle scattering \citep{Fu2022}.
A study of the p=1 case using this type of method \citep{Wetering2021} showed the azimuthal current drive efficiency at the low collision limit is surprisingly high, especially in comparison with LHCD or ECCD \citep{Fisch1978, Karney1979, Fisch1980, Fisch1981, Fisch1987}, which entail mechanisms that push current parallel to the magnetic field. There are a few noteworthy similarities
between the result reported in Ref.\,\citep{Wetering2021} and the
present findings, despite the fact that our analysis does not yet
include collisions. In Ref.\,\citep{Wetering2021}, while the electron
orbit is bounded without collisions, the high current drive efficiency
is correlated with the orbit radial expansion under the influence
of collisions. The expansion induced by collisions for otherwise stable
orbits is consistent with our conclusion that the case of $p=1$ is
structurally unstable, i.e., a small perturbation to the system parameter
will render the dynamics unstable. Furthermore, the connection between
expanded orbits and larger current drive efficiency agrees with our
finding that orbit instabilities significantly enhanced current drive.
These topics will be studied in detail as the next step. 
\begin{acknowledgments}
Eric Palmerduca was supported by Cornell NNSA 83228-10966 {[}Prime
No. DOE (NNSA) DE-NA0003764{]}. Hong Qin and Samuel Cohen were supported
by U.S. Department of Energy (DE-AC02-09CH11466). We thank Prof. Nathaniel
Fisch for fruitful discussions. This work is inspired by his original
contribution to this topic. 
\end{acknowledgments}

 \bibliographystyle{apsrev4-2}
\bibliography{bibliography}

\end{document}